\documentclass[onecolumn,draftcls]{IEEEtran}

\usepackage[latin1]{inputenc}
\usepackage[english]{babel}
\usepackage{amsmath,amsfonts,amssymb,dsfont}
\usepackage{stackrel,cite,balance}
\usepackage{graphicx,epstopdf,color}
\usepackage{pgfplots} 
\newcommand{\refE}[1]   {(\ref{eqn:#1})}
\newcommand{\refS}[1]   {Section~\ref{sec:#1}}
\newcommand{\refF}[1]   {Fig.~\ref{fig:#1}}
\newcommand{\refFig}[1] {Figure~\ref{fig:#1}}
\usepackage{notation}  
\newcommand{\Sph}[1]{\Sc_{#1}}
\newcommand{\Sphi}[1]{\Sc_{\text{i},#1}}
\newcommand{\Spho}[1]{\Sc_{\text{o},#1}}

\begin{document}
\title{The Error Probability of Generalized Perfect Codes
       via the Meta-Converse}
\author{Gonzalo~Vazquez-Vilar,~\IEEEmembership{Member,~IEEE,}
        Albert~Guill{\'e}n~i~F{\`a}bregas,~\IEEEmembership{Senior Member,~IEEE,}\\
        Sergio Verd\'u,~\IEEEmembership{Fellow,~IEEE}%
\thanks{This work has been funded in part by the European Research Council (ERC) under grants 714161 and 725411, by the Spanish Ministry of Economy and Competitiveness under Grants TEC2016-78434-C3 and IJCI-2015-27020, by the National Science Foundation under Grant CCF-1513915, by the National Science Foundation under Grant CCF-1513915 and by the Center for Science of Information, an NSF Science and Technology Center under Grant CCF-0939370.}
\thanks{This work was presented in part at the 2016 International Z\"urich Seminar on Communication and Information and at the 2018 IEEE International Symposium on Information Theory, Vail, Colorado.}
}
\maketitle
 
\begin{abstract}
We introduce a definition of perfect and quasi-perfect codes for symmetric channels parametrized by an auxiliary output distribution. This notion generalizes previous definitions of perfect and quasi-perfect codes and encompasses maximum distance separable codes. The error probability of these codes, whenever they exist, is shown to coincide with the estimate provided by the meta-converse lower bound. We illustrate how the proposed definition naturally extends to cover almost-lossless source-channel coding and lossy compression.
\end{abstract}

\section{Introduction}\label{sec:intro}
In the context of reliable communication, binary hypothesis testing has proved instrumental in the derivation of converse bounds to the error probability. Using this method, the sphere-packing bound on the channel coding reliability function was derived in \cite{Shannon67I} (see also \cite{Haroutunian68,Blahut74, sason2008, altug2014} for alternative derivations and refinements).
More recently, the meta-converse of Polyanskiy  {\em et al.} \cite[Th. 27]{Pol10} proved that a surrogate binary hypothesis test can be used to accurately lower-bound the error probability in the finite blocklength regime. The performance of binary hypothesis testing between distributions $P_0$ and $P_1$ (in absence of priors) is characterized by the trade-off $\alpha_{\beta}\bigl(P_0, P_1\bigr)$, where $\alpha$ denotes the smallest error under $P_0$ achievable by any test with error under $P_1$ at most~$\beta$ (we refer the reader to Section \ref{sec:HT} for a formal definition). Then, \cite[Th. 27]{Pol10} establishes the following lower bound on the error probability of a code $\Cc$ with cardinality $M$ used over a channel $\pyx$,
\begin{align}
  \Pe(\Cc)
&\geq \inf_{\px} \sup_{Q_Y} \left\{
                 \alpha_{\frac{1}{M}} \bigl(\px \times \pyx, \px \times Q_Y \bigr)\right\}. \label{eqn:intro-metaconverse}
\end{align}
This bound is usually referred to as the meta-converse bound, since several previous converse bounds in the literature can be derived from it via relaxation. 
Particularized for $n$-uses of a memoryless binary symmetric channel (BSC), the meta-converse bound recovers the sphere-packing bound for BSC channels~\cite[Eq. (5.8.19)]{Gall68} (see~\cite[Sec. III.H]{Pol10}). In this setting, the meta-converse bound \refE{intro-metaconverse} coincides with the exact error probability whenever perfect or quasi-perfect codes exist. In particular, a binary code is said to be \textit{perfect} if non-overlapping Hamming spheres of radius $t$ centered on the codewords exactly fill out the space. Similarly, a \textit{quasi-perfect} code is defined as a code in which Hamming spheres of radius $t$ centered on the codewords are non-overlapping and Hamming spheres of radius $t+1$ cover the space, possibly with overlaps. This definition coincides with that of \textit{sphere-packed codes} introduced by Gallager \cite[Sec.~5.8]{Gall68}. Since quasi-perfect codes attain the lower bound \refE{intro-metaconverse}, they achieve the minimum error probability in a BSC among all the codes with the same blocklength and rate.

In this work, we extend this result for general channels under certain symmetry conditions. We generalize the definition of perfect and quasi-perfect codes beyond Hamming distance, and show that these codes attain equality in~\refE{intro-metaconverse}. Therefore, generalized quasi-perfect codes achieve the minimum error probability among all the codes with the same blocklength and rate. As an example, we study a family of $q$-ary symmetric erasure channels and we show that maximum-distance separable (MDS) codes are generalized quasi-perfect for these channels. As a result, we obtain an alternative proof of the optimality of MDS codes for $q$-ary symmetric erasure channels. Extensions to almost-lossless source-channel coding and lossy compression under an excess-distortion constraint are discussed.

A tightened version of the meta-converse, derived for a fixed code, was shown to coincide with the exact error probability in \cite[Th. 1]{vazquez16}. In contrast to \cite{vazquez16}, in this paper we show that the bound \refE{intro-metaconverse}, which applies to every code of cardinality $M$, also yields the exact error probability in certain cases. 
In~\cite{Hamada2000}, Hamada also studied a generalization of perfect and quasi-perfect codes beyond Hamming distance. Using a variation of the Fano metric \cite[Eq.~(9.10)]{Fano61}, Hamada derived a lower bound to the channel coding error probability. Our definition of quasi-perfect codes includes~\cite[Def.~1]{Hamada2000} as a special case and recovers Hamada's condition for achieving minimum error probability~\cite[Th. 3]{Hamada2000}. Nevertheless, the class of generalized codes considered here is more general than that in~\cite{Hamada2000} and shows connections not previously treated in the literature.

The structure of the paper is as follows. In \refS{HT} we introduce the binary hypothesis testing framework and notation used in the rest of the paper. In \refS{generalized-quasi-perfect} we introduce the system model and show the optimality of the so-called generalized quasi-perfect codes.  A family of erasure channels is studied in detail under this formulation in \refS{erasure-error-channels} and the optimality of MDS codes is shown. Sections \ref{sec:jscc} and \ref{sec:lsc} extend the notion of generalized quasi-perfect codes to almost-lossless source-channel coding and lossy compression under maximum excess-distortion probability, respectively. \refS{discussion} closes the paper with some final remarks.

\section{Binary Hypothesis Testing}
\label{sec:HT}
Consider a non-Bayesian binary hypothesis test discriminating between distributions $P_0$ and $P_1$ defined over some discrete alphabet $\Zc$. Let $T(z)$ denote the probability of the test deciding hypothesis $0$ (corresponding to $P_0$) given an observation~$z$, $0 \leq T(z)\leq 1$. Then, $1-T(z)$ is the probability of deciding  hypothesis $1$ (i.e., $P_1$). Let $\pi_{j|i}$ denote the error probability of deciding $j$ when $i\neq j$ is the true hypothesis. More precisely, we define
\begin{align}
 \pi_{0|1}(T)
     &\triangleq \sum_{z} T(z) P_1(z),
 \label{eqn:bht-type1error}\\
 \pi_{1|0}(T) 
     &\triangleq 1- \sum_{z} T(z) P_0(z),
 \label{eqn:bht-type0error}
\end{align}
and we denote the minimum error probability $\pi_{1|0}$ among all tests $T$ with $\pi_{0|1}$ at most $\beta$, as
\begin{align}
\alpha_{\beta}\bigl(P_0, P_1\bigr)
  \triangleq \inf_{{T: \pi_{0|1}(T) \leq \beta}}  \pi_{1|0}(T).
 \label{eqn:bht-alpha}
\end{align}
Neyman and Pearson (NP) provided in \cite{NeyPea33} an explicit form for the test achieving the optimal trade-off $\alpha_{\beta}\bigl(P_0, P_1\bigr)$. In particular, for any $\gamma \geq 0$, $\theta \in [0,1]$, the NP test is given by
\begin{align}\label{eqn:HT-NP-test}
  T_{\text{NP}} (z) \triangleq \openone\biggl[\frac{P_0(z)}{P_1(z)} > \gamma\biggr] + \theta \openone\biggl[\frac{P_0(z)}{P_1(z)} = \gamma\biggr] ,
\end{align}
where $\openone[\cdot]$ denotes the indicator function.
$T_{\text{NP}}$ achieves the optimal trade-off $\alpha_{\beta}\bigl(P_0, P_1\bigr) = \pi_{1|0}(T_{\text{NP}})$ when $\gamma$ and $\theta$ are chosen such that $\beta = \pi_{0|1}(T_{\text{NP}})$. This result is usually known as NP Lemma. A direct consequence of the NP Lemma is the following alternative characterization of the optimal error probability trade-off $\alpha_{\beta}\bigl(P_0, P_1\bigr)$.
\begin{lemma}\label{lem:HT-im-formulation}
For any non-Bayesian binary hypothesis test discriminating between $P_0$ and $P_1$, 
\begin{align}
\alpha_{\beta}\bigl(P_0, P_1\bigr)
=\,  \sup_{\gamma \geq 0} \left\{ \PP\biggl[  \frac{P_0(Z_0)}{P_1(Z_0)} \leq \gamma\biggr]
      + \gamma \PP\biggl[ \frac{P_0(Z_1)}{P_1(Z_1)} > \gamma\biggr]
           - \gamma\beta \right\},\label{eqn:HT-im-formulation}
\end{align}
where $Z_i \sim P_i$, $i=0,1$.
\end{lemma}

\section{Generalized Perfect Codes}\label{sec:generalized-quasi-perfect}


An equiprobable message $m\in\{1,\dots,M\}$ is to be transmitted over a channel with transition probability $\pyx$, input $x\in \Xc$ and output $y\in \Yc$, and  where $\Xc$ and $\Yc$ are the one-shot input/output discrete alphabets. A channel code is the set of codewords $\Cc = \{x_1, \ldots, x_M\}$ $x_i\in\Xc$ for $i=1,\dotsc,M$, assigned to each of the messages. Under maximum likelihood (ML) decoding, the error probability for the code $\Cc$  is given by
\begin{align}
  \Pe(\Cc)
    &= 1 - \frac{1}{M} \sum_{y} \max_{x \in \Cc} \pyx(y|x).
    \label{eqn:cc-errorML}
\end{align}

We consider the following family of symmetric channels.
\begin{definition} \label{def:symmetric-DMC}
Let $F_{x}(\tau) \triangleq \PP \bigl[\pyx(Y|x)\geq \tau \bigr]$, where $Y\sim \pyxx$ and $ \tau \in [0,1]$. A channel $\pyx$ is \textit{symmetric} if $F_{x}(\tau)$ does not depend on the input $x$, 
\begin{align}
  F_{x}(\tau)  = F(\tau),\quad  \forall x\in\Xc,\quad  \tau \in [0,1]. 
\end{align}
\end{definition}
In the special case of discrete memoryless channels, Definition~\ref{def:symmetric-DMC} implies that the rows of the channel transition matrix (with inputs as rows and outputs as columns), $\pyx(\cdot | x)$, are permutations of each other. This definition coincides with that of uniformly dispersive channels of Massey \cite[Sec. 4.2]{Massey_ADIT1_notes} and is less restrictive than those of Cover and Thomas~\cite[Sec. 7.2]{Cover06} and Gallager~\cite[p. 94]{Gall68}. The definition in \cite[Sec. 7.2]{Cover06} additionally requires that the columns of the channel transition matrix are permutations of each other. The definition in \cite[p. 94]{Gall68} requires the channel transition matrix to be partitioned in submatrices such that each submatrix fulfills the conditions in \cite[Sec. 7.2]{Cover06}. Relations among these notions are investigated in \cite[Sec. VI.B]{Pol13}.

Let $\qy$ be an auxiliary distribution defined on the output alphabet $\Yc$. For an observation $y\in\Yc$, the codeword $x\in\Cc$ that maximizes the metric $\pyx(y|x)$ also maximizes the metric $q(x,y) = \frac{\pyx(y|x)}{\qy(y)}$. We conclude that the decoding regions induced by the ML decoder (with metric ${\pyx(y|x)}$) and those of the maximum metric decoder (with metric $q(x,y)$) coincide. This obvious fact proves to be instrumental next.

For any $\tau\geq0$ and any distribution $Q$ defined over $\Yc$, we define $\Sph{x}(\tau,Q)$ to be the set of outputs $y$ with likelihood given input $x$ at least $\tau \qy(y)$, i.e.,
\begin{align}\label{eqn:def-SQ}
  \Sph{x}(\tau,Q) \triangleq \left\{ y \in \Yc \,\Big|\, \tfrac{\pyx(y|x)}{\qy(y)} \geq \tau \right\}.
\end{align}
We  refer to $\Sph{x}(\tau,Q)$ as a sphere of radius $\tau$ centered on $x$, although in general $\Xc\neq\Yc$ and $q(x,y) \triangleq \frac{\pyx(y|x)}{\qy(y)}$ is not a distance measure. This metric is equivalent to the Fano metric \cite[Eq.~(9.10)]{Fano61}, defined as $-\log q(x,y) = \log\tfrac{\qy(y)}{\pyx(y|x)}$. For channels such as the BSC, $\log \pyx(y|x)$ is an affine function of the Hamming distance between $x$ and $y$ and, hence, $\Sph{x}(\tau,Q)$ becomes a Hamming sphere when $\qy$ is the equiprobable distribution. 

We define the interior and the outer shell of $\Sph{x}(\tau,Q)$ as 
\begin{align}
  \Sphi{x}(\tau,Q) \triangleq \left\{ y \in \Yc \,\Big|\, \tfrac{\pyx(y|x)}{\qy(y)} > \tau\right\},
\label{eqn:def-SQi}\\
  \Spho{x}(\tau,Q) \triangleq \left\{ y \in \Yc \,\Big|\, \tfrac{\pyx(y|x)}{\qy(y)} = \tau \right\}.
\label{eqn:def-SQo}
\end{align}

We consider the set of distributions $\qy$ such that the tilted channel $\pyxtilde (y|x)\propto \frac{\pyx(y|x)}{\qy(y)}$ remains symmetric. More precisely, we define the set of symmetry-preserving auxiliary output distributions
\begin{align}\label{eqn:def-Qc}
  \Qc \triangleq \Bigl\{ \qy\in \Pc(\Yc) \;\big|\; 
   F_{x}(\tau,Q)  = F(\tau,Q),\ \forall x\in\Xc,\ \tau \geq 0\Bigr\},
\end{align}
where $F_{x}(\tau,Q)  \triangleq \PP \bigl[Y\in\Sph{x}(\tau,Q)\bigr]$ with $Y\sim \pyxx$, and $\Pc(\Ac)$ denotes the set of all probability distributions over alphabet $\Ac$. That is, $\Qc$ corresponds to the set of auxiliary distributions $Q$ such that the probability of $\Sph{x}(\tau,Q)$ under $\pyxx$ is independent of $x$ for any $\tau \geq 0$.

For symmetric channels $\pyx$, the set $\Qc$ is non-empty as it always includes the equiprobable distribution, and it may include other auxiliary distributions.
For example, consider a single use of the binary erasure channel (BEC) with erasure symbol $\mathsf{e}$. In this case, any distribution of the form $\qy(0) = \qy(1) = \xi$, $\qy(\mathsf{e}) = 1 - 2\xi$, does not alter the symmetry of the original channel, and therefore it is included in $\Qc$. This example will be studied in detail in~\refS{erasure-error-channels}.

For a fixed $Q\in\Qc$, we use the short-hand notation $\QQ[\Ac] \triangleq \PP [ Y \in \Ac ], Y \sim Q$.
\begin{lemma}\label{lem:vol-Q-spheres}
Let $\pyx$ be a symmetric channel according to Definition \ref{def:symmetric-DMC} and $Q\in\Qc$ defined in \refE{def-Qc}.
Then, the probabilities $\QQ\bigl[ \Sph{x}(\tau,Q) \bigr]$, $\QQ\bigl[ \Sphi{x}(\tau,Q) \bigr]$ and $\QQ\bigl[ \Spho{x}(\tau,Q) \bigr]$ are independent of $x\in\Xc$ for any $\tau\geq 0$.
\end{lemma}
\begin{IEEEproof}
We  prove that the term $\QQ\bigl[ \Spho{x}(\tau,Q) \bigr]$ does not depend on $x$. Then, the independence of the other two terms follows since
\begin{align}
  \QQ\bigl[ \Sph{x}(\tau,Q) \bigr]
    &= \sum_{\tau' \in \Lc_{\qy},\, \tau'\geq\tau}\QQ\bigl[ \Spho{x}(\tau',Q) \bigr],\\
  \QQ\bigl[ \Sphi{x}(\tau,Q) \bigr]
    &= \sum_{\tau' \in \Lc_{\qy},\, \tau'>\tau} \QQ\bigl[ \Spho{x}(\tau',Q) \bigr],
\end{align}
where $\Lc_{\qy}$ is defined as
\begin{align}
  \Lc_{\qy} \triangleq \left\{ \tau\in\RR \,\Big|\,  \exists x\in\Xc, \exists y\in\Yc, \tfrac{\pyx(y|x)}{\qy(y)} = \tau \right\}. \label{eqn:def-LcQ}
\end{align}

To show that $\QQ\bigl[ \Spho{x}(\tau,Q) \bigr]$ is independent of $x$, we write
\begin{align}
  \QQ\bigl[ \Spho{x}(\tau,Q) \bigr]
  &= \sum\nolimits_{y} \qy(y)  \openone \left[ \pyx(y|x) = \tau \qy(y) \right]\\
  &= \frac{1}{\tau} \sum\nolimits_{y} \pyx(y|x) \openone \left[ \pyx(y|x) = \tau \qy(y) \right].
     \label{eqn:proof-vol-Q-spheres}
\end{align}
According to the definition of $\Qc$ in \refE{def-Qc}, for any $Q\in\Qc$,
\begin{align}
   F_{x}(\tau,Q)
     &=\sum\nolimits_{y} \pyx(y|x)  \openone \left[ \pyx(y|x) \geq \tau \qy(y) \right]
\end{align}
does not depend on the specific value of $x$ for any \mbox{$\tau\geq0$}. 
Then, noting that the summation in \refE{proof-vol-Q-spheres} is given by $\lim_{\delta \to 0} \left( F_{x}(\tau,Q) - F_{x}(\tau+\delta,Q) \right)$, the result follows.
\end{IEEEproof}

Then, according to Lemma~\ref{lem:vol-Q-spheres}, we define for symmetric channels the probability measures
\begin{align}
\mathsf{Q}(\tau)
    \triangleq \QQ\bigl[ \Sph{x}(\tau,Q) \bigr],\quad 
\mathsf{Q}_{\text{i}}(\tau)
    \triangleq \QQ\bigl[ \Sphi{x}(\tau,Q) \bigr], \quad
\mathsf{Q}_{\text{o}}(\tau)    
    \triangleq \QQ\bigl[ \Spho{x}(\tau,Q) \bigr].
    \label{eqn:Q-prob-def}
\end{align}


For a fixed code $\Cc$ and auxiliary distribution $\qy\in\Qc$, we let $\eta \geq 0$ be the largest value such that
$\medcup_{x\in\Cc} \Sph{x}(\eta,Q) = \Yc$. Similarly, let $\nu \geq 0$ be the smallest value such that the codeword centered sets $\bigl\{ \Sphi{x} (\nu,Q) \bigr\}_{x \in \Cc}$ are disjoint. We refer to $\eta$ and $\nu$ as the \textit{covering} and \textit{packing radius} of the code $\Cc$ with respect to $Q$, respectively. Intuitively, $\Sphi{x} (\nu,Q)$ is the largest sphere packed inside the ML decoding region corresponding to $x \in \Cc$. Similarly, $\Sph{x} (\eta,Q)$ is the smallest sphere centered at $x \in \Cc$ which completely covers the corresponding ML decoding region.

\begin{definition}\label{def:perfect-Q-code}
A code $\Cc$ is \textit{generalized perfect} for $\pyx$, if there exists $\gamma \geq 0$ and $\qy\in\Qc$ such that the codeword-centered sets $\bigl\{ \Sph{x} (\gamma,Q) \bigr\}_{x \in \Cc}$ are disjoint and
\begin{align}\label{eqn:cup-Q-codes}
 \bigcup_{x\in\Cc}\Sph{x}(\gamma,Q) = \Yc.
\end{align}
A code is \textit{generalized quasi-perfect} if  there exists $\gamma \geq 0$ and $\qy\in\Qc$ such that \refE{cup-Q-codes} is satisfied and the codeword-centered sets $\bigl\{ \Sphi{x} (\gamma,Q) \bigr\}_{x \in \Cc}$ are disjoint.\footnote{While the sets $\Sph{x}(\gamma,Q)$ and $\Sphi{x}(\gamma,Q)$ are a function of the parameters $\gamma$ and $Q$, they depend only on their product (see \refE{def-SQ} and \refE{def-SQi}). Therefore, the two parameters $\gamma \geq 0$ and $\qy\in\Qc$ appearing in the definition of generalized perfect and quasi-perfect codes could be replaced by a single unnormalized function $f(y) = \gamma Q(y)$.}
\end{definition}

Note that for generalized quasi-perfect codes the covering and packing radius coincide. The definition of quasi-perfect codes includes perfect codes as a special case. 
To avoid ambiguities, for perfect codes we require that $\gamma$ is the largest value satisfying~\refE{cup-Q-codes}. For this value of $\gamma$, \begin{align}\label{eqn:cup-Q-interior}
  \bigcup_{x\in\Cc} \Sphi{x} (\gamma,Q) \subset \Yc.
\end{align}


The main result in this work, Theorem~\ref{thm:cc-Q-optimality}, is a consequence of the following lemma, which is a refinement of \cite[(9.15)-(9.16)]{Fano61}.

\begin{lemma}\label{lem:cc-Q-error}
Let $\pyx$ be a symmetric channel  according to Definition \ref{def:symmetric-DMC} and let $\qy\in\Qc$. The error probability of any code $\Cc$ with $M$ codewords satisfies, for any $\gamma\geq 0$ and any $Q\in\Qc$,
\begin{align}
  \Pe(\Cc)
    &\geq \gamma \biggl(\mathsf{Q}_{\text{i}}(\gamma)-\frac{1}{M}\biggr) +\!\sum_{\tau \in \Lc_{\qy},\,\tau\leq\gamma}\!\tau \mathsf{Q}_{\text{o}}(\tau),\label{eqn:cc-Q-error}
\end{align}
where $\Lc_{\qy}$ is defined in \refE{def-LcQ}.
Furthermore, the lower bound \refE{cc-Q-error} holds with equality if and only if $\Cc$ is generalized quasi-perfect and $\gamma$ and $\qy$ are the parameters (not necessarily unique) satisfying the conditions in Definition~\ref{def:perfect-Q-code}.
\end{lemma}
\begin{IEEEproof}
Let $\Cc = \{x_1, \ldots, x_M\}$ be an arbitrary code. 
We consider a deterministic ML decoder which partitions the output space into disjoint decoding regions $\{\Dc_1, \ldots, \Dc_M\}$. The error probability \refE{cc-errorML} becomes
\begin{align}
  \Pe(\Cc)
    &= 1 - \frac{1}{M} \sum_{m=1}^{M} \sum_{y\in\Dc_m} \pyx(y|x_m).
    \label{eqn:cc-Q-epsC-0}
\end{align}

For an observed $y$, the codeword $x\in\Cc$ that maximizes the metric $\pyx(y|x)$ coincides with the one maximizing the metric $q(x,y) =\frac{\pyx(y|x)}{\qy(y)}$. 
Then, using the definition of the covering and packing radius $\eta$ and $\nu$, respectively, it follows that
\begin{align}
  \Sphi{x_m}(\nu,Q) \subseteq \Dc_m \subseteq \Sph{x_m}(\eta,Q),
\end{align}
for $1 \leq m \leq M$. As a result, $\Dc_m$ can be decomposed as
\begin{align}
  \Dc_m &= \Sphi{x_m}(\nu,Q) 
   \stackrel[\substack{\tau \in \Lc_Q,\\ \eta\leq\tau\leq\nu}]{}{\medcup} \bigl(\Dc_m \cap \Spho{x_m}(\tau,Q)\bigr), \label{eqn:DmQ-decomposition}
\end{align}
and \refE{cc-Q-epsC-0} becomes
\begin{align}
  \Pe(\Cc)
    = 1 &- \frac{1}{M} \sum_{m=1}^{M} \Biggl(\,\sum_{y\in\Sphi{x_m}(\nu,Q)} \hspace{-2mm}\pyx(y|x_m)
  + \sum_{\substack{\tau \in \Lc_Q,\\ \eta\leq\tau\leq\nu}} \sum_{y\in\{\Dc_m \cap \Spho{x_m}(\tau,Q)\}} \hspace{-2mm} \pyx(y|x_m) \Biggr).
    \label{eqn:cc-Q-epsC-1}
\end{align}
Since $\frac{\pyx(y|x)}{\qy(y)}=\tau$ for any $y\in\Spho{x}(\tau,Q)$, we write
\begin{align}
\sum_{y\in\Sphi{x}(\nu)} \pyx(y|x)
&= \sum_{y\in\Sphi{x}(\nu,Q)}  \frac{\pyx(y|x)}{\qy(y)} \qy(y)\label{eqn:cc-Q-sumSph-0}\\
 &= \sum_{\tau \in \Lc_Q, \tau>\nu} \; \sum_{y\in\Spho{x}(\tau,Q)}  \tau \qy(y) \label{eqn:cc-Q-sumSph-1}\\
&= \sum_{\tau \in \Lc_Q, \tau>\nu} 
 \tau \mathsf{Q}_{\text{o}}(\tau), \label{eqn:cc-Q-sumSph-2}
\end{align}
where in~\refE{cc-Q-sumSph-2} we used Lemma \ref{lem:vol-Q-spheres} and $\mathsf{Q}_{\text{o}}(\tau) = \QQ\bigl[ \Spho{x}(\tau,Q) \bigr]$ as defined in~\refE{Q-prob-def}. Similarly, 
\begin{align}
\!\sum_{y\in\{\Dc_m \cap \Spho{x}(\tau,Q)\}}\!\pyx(y|x)
 &=\!\sum_{y\in\{\Dc_m \cap \Spho{x}(\tau,Q)\}}\!\tau \qy(y)\\
&= \tau \mathsf{Q}_{\text{o},m}(\tau). \label{eqn:cc-Q-sumSph-3}
\end{align}
where we abbreviate $\mathsf{Q}_{\text{o},m}(\tau) \triangleq \QQ\bigl[\Dc_m \cap \Spho{x_m}(\tau,Q)\bigr]$.

Substituting \refE{cc-Q-sumSph-2} and \refE{cc-Q-sumSph-3} in \refE{cc-Q-epsC-1}, yields 
\begin{align}
\!\Pe(\Cc) &= 1\!-\!\Biggl( \sum_{\substack{\tau \in \Lc_Q,\\ \tau>\nu}}\!\tau \mathsf{Q}_{\text{o}}(\tau)  + \frac{1}{M} \sum_{m=1}^M\!\sum_{\substack{\tau \in \Lc_Q,\\ \eta\leq\tau\leq\nu}}\!\tau \mathsf{Q}_{\text{o},m}(\tau)\!\Biggr).\!\label{eqn:cc-Q-epsC-2}
\end{align}

Since $\{ \Dc_m\}_{m=1}^M$ defines a partition of the output space, $\sum_{m=1}^{M} \QQ\bigl[\Dc_m\bigr]=1$. Using \refE{DmQ-decomposition} and the definitions of $\mathsf{Q}_{\text{i}}(\cdot)$ and $\mathsf{Q}_{\text{o},m}(\cdot)$, we obtain
\begin{align}
1 &= \sum_{m=1}^{M} \QQ\bigl[\Dc_m\bigr]
  = M \mathsf{Q}_{\text{i}}(\nu) + \sum_{m=1}^M \sum_{\substack{\tau \in \Lc_Q,\\ \eta\leq\tau\leq\nu}} \mathsf{Q}_{\text{o},m}(\tau).
  \label{eqn:cc-Q-space-partition-0}
\end{align}
Upon rearranging terms,~\refE{cc-Q-space-partition-0} yields
\begin{align}
\nu \left(
\frac{1}{M} - \mathsf{Q}_{\text{i}}(\nu)\right) &=  \frac{1}{M} \sum_{m=1}^M \sum_{\substack{\tau \in \Lc_Q,\\ \eta\leq\tau\leq\nu}} \nu\mathsf{Q}_{\text{o},m}(\tau)\label{eqn:cc-Q-space-partition-1}\\
        &\geq  \frac{1}{M} \sum_{m=1}^M \sum_{\substack{\tau \in \Lc_Q,\\ \eta\leq\tau\leq\nu}} \tau \mathsf{Q}_{\text{o},m}(\tau).\label{eqn:cc-Q-space-partition-2}
\end{align}
Then, using \refE{cc-Q-space-partition-1}-\refE{cc-Q-space-partition-2} in \refE{cc-Q-epsC-2}, it follows that  $\Pe(\Cc) \geq \Gamma(\nu)$ where
\begin{align}
\!\Gamma(\nu) \triangleq 1\!-\!\Biggl( \sum_{\tau \in \Lc_Q, \tau>\nu} \tau \mathsf{Q}_{\text{o}}(\tau)  + \nu \left(
\frac{1}{M} - \mathsf{Q}_{\text{i}}(\nu)\right)\!\Biggr).\!\label{eqn:cc-Q-epsC-3}
\end{align}
For quasi-perfect codes satisfying Definition~\ref{def:perfect-Q-code}, there exist $\qy\in\Qc$ and $\gamma = \nu = \eta$ such that covering and packing radius coincide. Then, for this choice of parameters, the inequality~\refE{cc-Q-space-partition-2} becomes equality and $\Pe(\Cc) = \Gamma(\gamma)$. We conclude that, for a generalized quasi-perfect code $\Cc$, \refE{cc-Q-error} holds with equality for any choice (not necessarily unique) of $\gamma$ and $\qy$ satisfying the conditions in Definition~\ref{def:perfect-Q-code}.

If $\Cc$ is not generalized quasi-perfect, $\nu > \eta$ for every $\qy\in\Qc$ and the inequality \refE{cc-Q-space-partition-2} is strict. Then, $\Pe(\Cc) > \Gamma(\nu)$. For any choice of $\gamma\geq 0$ not necessarily equal to the packing radius $\nu$, we next show that $\Pe(\Cc) > \Gamma(\gamma)$. First, note that for $\gamma > \nu$, both \refE{cc-Q-epsC-2} and \refE{cc-Q-space-partition-1}-\refE{cc-Q-space-partition-2} still hold substituting $\nu$ by $\gamma$. Then, the discussion above still applies.

Assume now that $\eta\leq\gamma < \nu$.
We can rewrite \refE{cc-Q-epsC-2} as
\begin{align}
\!\Pe(\Cc) = 1&-\Biggl( \sum_{\substack{\tau \in \Lc_Q,\\ \tau>\gamma}}\!\tau \mathsf{Q}_{\text{o}}(\tau)  + \frac{1}{M} \sum_{m=1}^M\!\sum_{\substack{\tau \in \Lc_Q,\\ \eta\leq\tau\leq\gamma}}\!\tau \mathsf{Q}_{\text{o},m}(\tau)\Biggr)
+ \frac{1}{M} \sum_{m=1}^M\!\sum_{\substack{\tau \in \Lc_Q,\\ \gamma <\tau\leq\nu}}\!\tau \Delta_{m}(\tau).\!\label{eqn:cc-Q-epsC-2-bis}
\end{align}
where $\Delta_{m}(\tau)\triangleq \mathsf{Q}_{\text{o}}(\tau)-\mathsf{Q}_{\text{o},m}(\tau) $.
Similarly, \refE{cc-Q-space-partition-0} becomes
\begin{align}
\!1 &= M \mathsf{Q}_{\text{i}}(\gamma) +\! \sum_{m=1}^M \sum_{\substack{\tau \in \Lc_Q,\\ \eta\leq\tau\leq\gamma}}\hspace{-1.5mm}\mathsf{Q}_{\text{o},m}(\tau)
- \sum_{m=1}^M \sum_{\substack{\tau \in \Lc_Q,\\ \gamma<\tau\leq\nu}}\hspace{-1.5mm} \Delta_{m}(\tau).\!
\end{align}
Following analogous steps as in 
\refE{cc-Q-space-partition-1}-\refE{cc-Q-space-partition-2}, via \refE{cc-Q-epsC-2-bis} we obtain
\begin{align}
\!\Pe(\Cc) &\geq \Gamma(\gamma) + \frac{1}{M} \sum_{m=1}^M \sum_{\substack{\tau \in \Lc_Q,\\ \gamma <\tau\leq\nu}} (\tau-\gamma) \Delta_{m}(\tau).\label{eqn:cc-Q-epsC-2-bound}
\end{align}
All terms in the inner sum in \refE{cc-Q-epsC-2-bound} satisfy $\tau-\gamma > 0$ and $\Delta_{m}(\tau) \geq 0$. If the code $\Cc$ is not generalized quasi-perfect, then, either $\Pe(\Cc) > \Gamma(\gamma)$ or $\Delta_{m}(\tau) > 0$ for at least one term in the sum. As the same proof steps follow for $\gamma < \eta$, we conclude that $\Pe(\Cc) > \Gamma(\gamma)$ for any $\gamma \geq 0$, $Q\in\Qc$, provided that $\Cc$ is not quasi-perfect.
\end{IEEEproof}

We are now ready to state the main result of this section, which shows that the ML decoding error probability of generalized perfect and quasi-perfect codes coincides with the meta-converse lower bound \refE{intro-metaconverse}.
\begin{theorem}\label{thm:cc-Q-optimality}
Let $\pyx$ be a symmetric channel according to Definition \ref{def:symmetric-DMC} and $\Cc$ be generalized quasi-perfect code according to Definition \ref{def:perfect-Q-code}. Then, $\Cc$ attains the minimum error probability among all codes with $M$ codewords, and it is given by
\begin{align}
  \Pe(\Cc)
    &=  \inf_{\px} \max_{\qy\in\Qc} \left\{
                 \alpha_{\frac{1}{M}} \bigl(\px \times \pyx, \px \times \qy \bigr)\right\}\label{eqn:cc-Q-optimality}\\
    &=  \max_{\qy\in\Qc} \left\{ \alpha_{\frac{1}{M}} \bigl(\pyxx, \qy \bigr)\right\}, \text{ for every } x\in\Xc.\label{eqn:cc-Q-optimality-bis}
\end{align}
Conversely, any code for which \refE{cc-Q-optimality}-\refE{cc-Q-optimality-bis} hold is generalized quasi-perfect.
\end{theorem}
\begin{IEEEproof}
Let us consider the hypothesis test in \refE{cc-Q-optimality}. We  apply Lemma~\ref{lem:HT-im-formulation} with $P_0 \leftarrow \px \times \pyx$ and $P_1 \leftarrow \px\times\qy$.
Using the definition of the set $\Sphi{x}(\cdot)$ and $\mathsf{Q}_{\text{i}}(\cdot)$ in Lemma~\ref{lem:vol-Q-spheres} yields
\begin{align}
\alpha_{\frac{1}{M}}\bigl(\px \times \pyx, \px\times\qy\bigr)
&=\sup_{\gamma \geq 0} \Biggl\{ \sum_{x,y\notin \Sphi{x}(\gamma,Q)}\!\px(x)\pyx(y|x)
\!+\!\gamma \mathsf{Q}_{\text{i}}(\gamma)
\!-\!\frac{\gamma}{M}\!\Biggr\}.\!\label{eqn:cc-Q-alpha-1}
\end{align}
For any $y\in \Spho{x}(\tau,Q),\tau \in \Lc_{\qy}$, where $\Lc_{\qy}$ is defined in \refE{def-LcQ}, it holds that $\frac{\pyx(y|x)}{\qy(y)}=\tau$. Then,
\begin{align}
\sum_{y\notin \Sphi{x}(\gamma,Q)} \pyx(y|x)
   &=  \sum_{\substack{{\tau \in \Lc_{\qy},\,\tau\leq\gamma,}\\{y\in \Spho{x}(\tau,Q)}}} \frac{\pyx(y|x)}{\qy(y)} \qy(y)\\
 &= \sum_{\tau \in \Lc_{\qy},\,\tau\leq\gamma}\sum_{y\in \Spho{x}(\tau,Q)} \tau \qy(y)\\
 &= \sum_{\tau \in \Lc_{\qy},\,\tau\leq\gamma}\tau \mathsf{Q}_{\text{o}}(\tau),
\end{align}
which does not depend on $x$ (see Lemma~\ref{lem:vol-Q-spheres}).
Then, \refE{cc-Q-alpha-1} becomes
\begin{align}
 \alpha_{\frac{1}{M}}\bigl(\px \times \pyx, \px\times\qy\bigr)
=\max_{\gamma \geq 0} \Biggl\{ \sum_{\tau \in \Lc_{\qy},\,\tau\leq\gamma}\tau \mathsf{Q}_{\text{o}}(\tau)
+ \gamma \mathsf{Q}_{\text{i}}(\gamma)
- \tfrac{\gamma}{M} \Biggr\}.\!\label{eqn:cc-Q-alpha-2}
\end{align}

According to \refE{intro-metaconverse}, the right-hand side of \refE{cc-Q-alpha-2} is a lower bound to $\Pe(\Cc)$. According to Lemma~\ref{lem:cc-Q-error}, the term in braces in \refE{cc-Q-alpha-2} is precisely the error probability of a generalized quasi-perfect code with parameters $\qy$ and $\gamma$. Then, whenever this code exists the lower bound \refE{cc-Q-alpha-2} is achievable and \refE{cc-Q-optimality} holds with equality. Moreover, \refE{cc-Q-optimality-bis} holds  since \refE{cc-Q-alpha-2} does not depend on $\px$ for symmetric channels and $\qy\in\Qc$.

Let now $\qy\in\Qc$ achieve \refE{cc-Q-optimality}-\refE{cc-Q-optimality-bis},
and let $\gamma$ be the maximizer in \refE{cc-Q-alpha-2}. It follows from Lemma~\ref{lem:cc-Q-error} that the term in braces in \refE{cc-Q-alpha-2} is the error probability of a code $\Cc$ if and only if $\Cc$ is generalized quasi-perfect and the parameters $\gamma$ and $\qy$ satisfy the conditions in Definition \ref{def:perfect-Q-code}. We conclude that, if \refE{cc-Q-optimality}-\refE{cc-Q-optimality-bis} hold, $\Cc$ must be generalized quasi-perfect.
\end{IEEEproof}


For any codebook $\Cc = \{x_1, \ldots, x_M\}$, we let $\pxC$ denote the distribution induced by $\Cc$, i.~e.,
$\pxC(x) \triangleq \frac{1}{M} \sum_{m=1}^M \mathds{1}\{ x = x_m \}$. It has been shown in \cite[Th. 1]{vazquez16} that the error probability of any code can be expressed as
\begin{align}
  \Pe(\Cc)
    &= \max_{\qy} \left\{
          \alpha_{\frac{1}{M}} \Bigl(\pxC \times \pyx,
                  \pxC \times \qy \Bigr) \right\}    \label{eqn:tight-meta}\\
    &\geq \min_{\px} \max_{\qy} \left\{
                 \alpha_{\frac{1}{M}} \bigl(\px \times \pyx, \px \times \qy \bigr)\right\}, \label{eqn:relaxed-meta}
\end{align}
Eq. \refE{tight-meta} shows that the meta-converse bound, when applied to a fixed code $\Cc$, coincides with the exact error probability $\Pe(\Cc)$. Theorem~\ref{thm:cc-Q-optimality} shows that, under certain symmetry conditions, the relaxation \refE{relaxed-meta} also coincides with the exact error probability, provided that a quasi-perfect code of cadinality $M$ exists for this channel.
Also, Theorem~\ref{thm:cc-Q-optimality} is more general than the result obtained by Hamada in \cite[Th. 3]{Hamada2000}. For instance, our result can be used to prove the optimality of MDS codes in $q$-ary erasure channels, as we show in the next section.

\section{Symmetric Erasure/Error Channels}\label{sec:erasure-error-channels}
Consider a symmetric erasure channel $\pyx$ with discrete input alphabet $\Xc$, $|\Xc|=q$, and output alphabet $\Yc= \Xc \cup \{\mathsf{e}\}$ where $\mathsf{e}$ corresponds to the erasure symbol:
\begin{align}
  \pyx(y|x) =
  \begin{cases}
    1-\delta-\eps, &y=x,\\
    \delta,        &y=\mathsf{e},\\
    \frac{\eps}{q-1},          &\text{otherwise}.
  \end{cases}\label{eqn:erasure-pyx}
\end{align}
When $q=2$, this channel includes as particular cases the BSC and the BEC with $\delta = 0$ and $\eps = 0$, respectively.

We consider $n$ uses of this channel. Let $\x=(x_1,\ldots,x_n)$ and $\y=(y_1,\ldots,y_n)$ denote the channel input and output, respectively. For a given pair of $\x$ and $\y$, we define the number of erasures and the number of flip-errors, respectively, as
\begin{align}
 e_{\y} &\triangleq \sum_{i} \openone[y_i = \mathsf{e}],\\
 d_{\x,\y} &\triangleq \sum_{i} \openone[x_i \neq y_i] - e_{\y}.
 \end{align}
The $n$-dimensional channel transition probability is given by
\begin{align}
\Pyx(\y|\x) &= \delta^{e_{\y}} \bigl(\tfrac{\eps}{q-1}\bigr)^{d_{\x,\y}} (1-\delta-\eps)^{n-e_{\y}-d_{\x,\y}}.\label{eqn:erasure-Pyx}
\end{align}
We  assume that  $\tfrac{\eps}{q-1} < 1-\delta-\epsilon$. Otherwise, observing the transmitted symbol at the output of the channel would be less likely than observing any of the other $q-1$ symbols. Particularized to the BSC (with $q=2$, $\delta=0$), this assumption boils down to the crossover probability being $\eps<\tfrac{1}{2}$.

We define the auxiliary distribution 
\begin{align}
  \Qy^{\star}(\y) &\triangleq
    \tfrac{1}{c}\delta^{e_{\y}} \bigl(\tfrac{\eps}{q-1}\bigr)^{\Psi(e_{\y})} (1-\delta-\epsilon)^{n-e_{\y}-\Psi(e_{\y})},\label{eqn:erasure-Qy}
\end{align}
where $c$ is a normalizing constant, and $\Psi(e)\geq 0$ is an arbitrary function of the number of erasures $e$, which can be optimized over. Intuitively, $\Psi(e)$ corresponds to the average number of flip-errors that a good code can correct when the output sequence is affected by $e$ erasures. For binary-input channels, a good choice for $\Psi(e)$ is given by
\begin{align}
   \Psi(e) = \max\left(0, \Bigl\lfloor\tfrac{\lceil n-\log_2 M\rceil - e + 1}{2}\Bigr\rfloor \right).\label{eqn:binary-De}
\end{align}

Since $\Qy^{\star}(\y)$ only depends on $\y$ via the number of erasures $e_{\y}$ it does not affect the symmetry of the vector channel $\Pyx$ and thus $\Qy^{\star} \in \Qc$. 
Theorem \ref{thm:cc-Q-optimality} is applied to this channel and auxiliary distribution $\qy=\Qy^{\star}$ to obtain the following result.

\begin{corollary}\label{cor:erasure-Q-optimality}
The error probability of any code $\Cc$ with cardinality $M$ used over the channel \refE{erasure-Pyx} satisfies
\begin{align}
\Pe(\Cc) \geq \sum_{e=0}^{n} \sum_{d=0}^{n-e} {n \choose e} {n-e \choose d} (q-1)^d \delta^{e} (1-\delta-\eps)^{n-e}
\left(\varphi^{\max(d,\Psi(e))} - \frac{\varphi^{\Psi(e)}}{M}\right)\!,\label{eqn:erasure-Q-optimality}
\end{align}
where $\varphi \triangleq \tfrac{\eps}{q-1} (1-\delta-\eps)^{-1}$ and $\Psi(e)\geq 0$ is any positive function of the number of erasures $e$. Moreover, if $\Cc$ is a generalized quasi-perfect code that satisfies Definition~\ref{def:perfect-Q-code} with $\gamma=c$ and $Q=\Qy^{\star}$ then  \refE{erasure-Q-optimality} holds with equality.
\end{corollary}
\begin{IEEEproof}
Let us consider the lower bound that follows from \refE{cc-Q-alpha-2} by fixing $\qy = \Qy^{\star}$ defined in \refE{erasure-Qy} and fixing $\gamma' = c$ to be the normalization factor appearing in~\refE{erasure-Qy},
\begin{align}
\max_{\qy\in\Qc} \Bigl\{ \alpha_{\frac{1}{M}}\bigl(\Px \times \Pyx, \Px\times\qy\bigr)\Bigr\}
&\geq \sum_{\x}\Px(\x) \bigggl(\sum_{\y\notin \Sphi{\x}(c,\Qy^{\star})} \Pyx(\y|\x)
 - c \bigggl( \frac{1}{M}- \sum_{\y\in \Sphi{\x}(c,\Qy^{\star})} \Qy^{\star}(\y) \bigggr)\! \bigggr).\label{eqn:MDS-alpha-1}
\end{align}
For $\qy=\Qy^{\star}$ defined in \refE{erasure-Qy},
the sets $\Sphi{\x}(c,\qy)$ can be expressed as
\begin{align}
  \Sphi{\x}(c,\qy) &= \left\{ \y \in \Yc \,\Big|\, 
    d_{\x,\y} < \Psi(e_\y)\right\}.\label{eqn:MDS-SphQi}
\end{align}

We  parametrize each output sequence $\y$ by the indices $e = e_{\y} \in [0,n]$ and $d = d_{\x,\y} \in [0,n-e_{\y}]$.  For a given $\x$, there are exactly ${n \choose e}{n-e \choose d} (q-1)^d$ output sequences $\y$ with indices $e,d$. Using this parametric representation, using \refE{MDS-SphQi} and the definitions of $\Pyx$ in \refE{erasure-Pyx} and $\Qy^{\star}$ in \refE{erasure-Qy}, we obtain
\begin{align}
\sum_{\y \notin \Sphi{\x}(c,\Qy^{\star})} \Pyx(\y|\x)
&= \sum_{e=0}^{n} \sum_{d=\Psi(e)}^{n-e} {n \choose e} {n-e \choose d} (q-1)^d (1-\delta-\eps)^{n-e}  \delta^{e} \varphi^{d},\\
\sum_{\y \in \Sphi{\x}(c,\Qy^{\star})} \Qy^{\star}(\y)
&= \frac{1}{c} \sum_{e=0}^{n}  \sum_{d=0}^{\Psi(e)-1} {n \choose e} {n-e \choose d} (q-1)^d (1-\delta-\eps)^{n-e}  \delta^{e} \varphi^{\Psi(e)}.
\end{align}
Substituting these expressions in \refE{MDS-alpha-1}, reorganizing terms, yields
\begin{align}
  \max_{\qy\in\Qc} \Bigl\{ \alpha_{\frac{1}{M}}\bigl(\Px \times \Pyx, \Px\times\qy\bigr)\Bigr\}
    &\geq \sum_{e=0}^{n} \sum_{d=0}^{n-e} {n \choose e} {n-e \choose d} (q-1)^d (1-\delta-\eps)^{n-e}  \delta^{e} \varphi^{\max(d,\Psi(e))} - \frac{c}{M}.\label{eqn:MDS-alpha-2}
\end{align}
Finally, noting that
\begin{align}
  \sum_{d=0}^{n-e} {n-e \choose d} (q-1)^d = q^{n-e},
\end{align}
we obtain for the normalizing constant in~\refE{erasure-Qy},
\begin{align}
c 
&= \sum_{e=0}^{n} {n \choose e} q^{n-e} (1-\delta-\eps)^{n-e}  \delta^{e} \varphi^{\Psi(e)}\label{eqn:MDS-c}
\end{align}
Substituting \refE{MDS-c} in \refE{MDS-alpha-2}, via the meta-converse bound \refE{intro-metaconverse}, we obtain \refE{erasure-Q-optimality}. 
According to Lemma \ref{lem:cc-Q-error}, this bound holds with equality if $\Cc$ is generalized $Q$-quasi-perfect with parameters $\gamma=c$ and $Q=\Qy^{\star}$. 
\end{IEEEproof}

Let $d_{\min}$ denote the minimum Hamming distance between any pair of codewords in $\Cc$. The Singleton bound~\cite[Th. 4.5.6]{Roman-1992} establishes the maximum number of codewords $M$ in a $q$-ary block code $\Cc$ of length $n$ and minimum distance~$d_{\min}$,
\begin{align}
  \log_q M \leq n-d_{\min}+1.\label{eqn:singleton}
\end{align}
Those codes achieving the Singleton bound with equality are termed  MDS codes. Examples of MDS codes include those that have only two complementary codewords thus having $d_{\min}=n$, non-redundant codes, i.e., $\Cc=\Xc$, for which $d_{\min}=1$, codes with a single parity symbol for which $d_{\min}=2$ and their corresponding dual codes. These are often called trivial MDS codes. In the case of binary alphabets, only trivial MDS codes exist. For non-binary alphabets, Reed-Solomon codes are an example of non-trivial MDS codes.

MDS codes are indeed generalized quasi-perfect codes for the $q$-ary erasure channel, given by $\Pyx$ in \refE{erasure-Pyx} when $\eps=0$. We note that $\lim_{\eps\to 0} \eps^{A} = 0$ for any $A>0$ and $\lim_{\eps\to 0} \eps^{A} = 1$ for $A=0$. Then, for any function $\Psi(e) \geq 0$ such that $\Psi(e)=0$ if, and only if $e > n-\log_q M$, \refE{erasure-Qy} becomes
\begin{align}
  \Qy^{\star}(\y) =
  \begin{cases}
    0,& e_{\y} \leq n-\log_q M,\\
    \tfrac{1}{c} \delta^{e_{\y}}(1-\delta)^{n-e_{\y}},& e_{\y} >  n-\log_q M.
  \end{cases}\label{eqn:erasure-Qy-bis}
\end{align}

Consider a generalized quasi-perfect code according to Definition~\ref{def:perfect-Q-code} with parameters $Q=\Qy^{\star}$ and $\gamma=c$ as defined in \refE{erasure-Qy-bis}. For the sets $\Sph{x}(\cdot)$ we  use the
convention that, when $\Qy^{\star}(\y)=0$,
\begin{align}
  \frac{\Pyx(\y|\x)}{\Qy^{\star}(\y)}=
  \begin{cases} 
    0,& \text{if } \Pyx(\y|\x)=0,\\
    \infty,& \text{if } \Pyx(\y|\x)>0.
  \end{cases}
\end{align}

The spheres induced by this code are such that their interior $\Sphi{\x}(c,\Qy^{\star})$ is the set of the output sequences $\y$ that are compatible with the input $\x$ with a number of erasures $e_{\y} \leq n-\log_q M$. Since the codeword-centered interiors do not overlap, the minimum distance of the code is at least $\lfloor n-\log_q M\rfloor +1$.
Since the codeword centered shells $\Spho{\x}(c,\Qy^{\star})$ overlap at some point, $d_{\min}$ is exactly 
\begin{align}
d_{\min} = \lfloor n-\log_q M\rfloor +1.
\end{align}
When $\log_q M$ is an integer, this expression coincides with the Singleton bound \refE{singleton}. As a result, we conclude that MDS codes are also quasi-perfect. By letting $\eps\to 0$  in  Corollary \ref{cor:erasure-Q-optimality} for any $\Psi(e)$ such that $\Psi(e)=0$ iff $e > n-\log_q M$, we obtain the following result.
\begin{corollary}\label{cor:MDS-optimality}
The error probability of any code $\Cc$ with cardinality $M$ used over a $q$-ary erasure channel satisfies
\begin{align}
\!\Pe(\Cc) &\geq\!\sum_{e= \lfloor n-\log_q M\rfloor+1}^{n}\!{n \choose e} \delta^{e} (1-\delta)^{n-e} \left(1 - \tfrac{q^{n-e}}{M}\right)\!,\!\label{eqn:MDS-optimality}
\end{align}
with equality if $\Cc$ is generalized quasi-perfect with parameters $\gamma=c$ and $Q=\Qy^{\star}$, as defined in \refE{erasure-Qy-bis}.
\end{corollary}

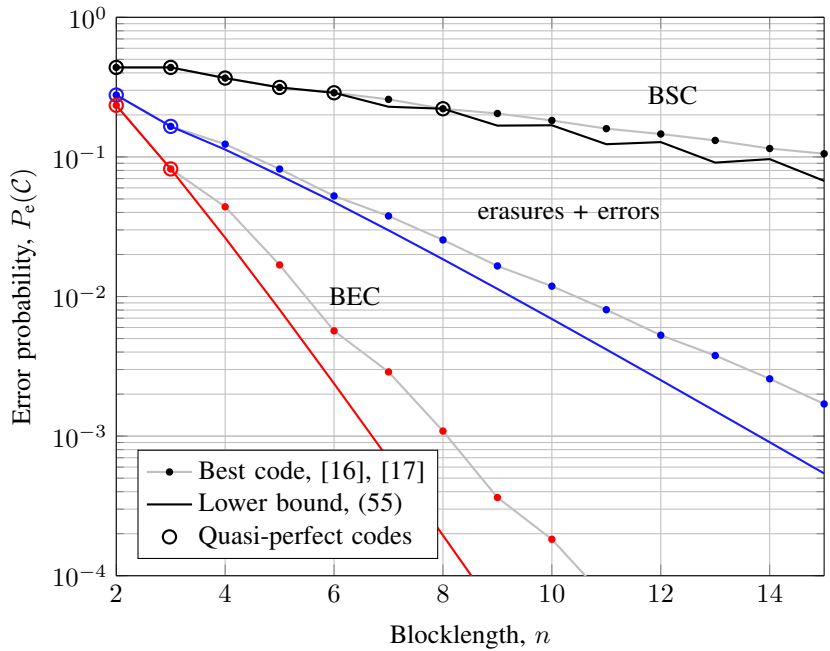
\begin{figure}[t]%
\centering
%
%
\begin{tikzpicture}

\begin{axis}[%
width=0.57\textwidth,
height=0.45\textwidth,
at={(0\textwidth,0\textwidth)},
scale only axis,
separate axis lines,
every outer x axis line/.append style={black},
every x tick label/.append style={font=\color{black}},
xmin=2,
xmax=15,
xlabel={Blocklength, $n$},
xmajorgrids,
every outer y axis line/.append style={black},
every y tick label/.append style={font=\color{black}},
ymode=log,
ymin=0.0001,
ymax=1,
yminorticks=true,
ylabel={Error probability, $\Pe(\Cc)$},
ymajorgrids,
yminorgrids,
legend style={at={(0.03,0.03)},anchor=south west, legend cell align=left,align=left,draw=black}
]
\addplot [color=white!50!black!50,line width=0.8pt,mark=*,mark options={black, scale=0.5, solid}]
  table[row sep=crcr]{%
2	0.4375\\
3	0.4375\\
4	0.3671875\\
5	0.314453125\\
6	0.2880859375\\
7	0.2584228515625\\
8	0.221343994140625\\
9	0.204658508300781\\
10	0.182411193847656\\
11	0.159468650817871\\
12	0.145911693572998\\
13	0.131246715784073\\
14	0.114870823919773\\
15	0.105399692431092\\
};
\addlegendentry{Best code,~\cite{moser2013,moser2017}};

\addplot [color=black,line width=0.8pt]
  table[row sep=crcr]{%
2	0.4375\\
3	0.4375\\
4	0.3671875\\
5	0.314453125\\
6	0.2880859375\\
7	0.228759765625\\
8	0.221343994140625\\
9	0.167579650878906\\
10	0.168506622314453\\
11	0.123316764831543\\
12	0.127661943435669\\
13	0.0910320878028872\\
14	0.0964418351650239\\
15	0.0673562958836557\\
};
\addlegendentry{Lower bound,~\refE{erasure-Q-optimality}};

\addplot [color=black,line width=0.8pt,only marks,mark=o,mark options={solid,scale=1.25}]
  table[row sep=crcr]{%
2	0.4375\\
3	0.4375\\
4	0.3671875\\
5	0.314453125\\
6	0.2880859375\\
8	0.221343994140625\\
};
\addlegendentry{Quasi-perfect codes};

\addplot [color=white!50!black!50,line width=0.8pt,mark=*,mark options={blue, scale=0.5, solid}]
  table[row sep=crcr]{%
2	0.2775\\
3	0.1655\\
4	0.1234125\\
5	0.0818543750000001\\
6	0.0525427499999996\\
7	0.0377532921875007\\
8	0.0254261439843757\\
9	0.0165547512968806\\
10	0.0118559819156274\\
11	0.00804283338870011\\
12	0.00528296577322295\\
13	0.00377805999127903\\
14	0.00257309978588831\\
15	0.00169957383723702\\
};

\addplot [color=blue!90!white,line width=0.8pt]  table[row sep=crcr]{%
2	0.2775\\
3	0.1655\\
4	0.1129125\\
5	0.073716875\\
6	0.0474436875\\
7	0.029824971875\\
8	0.018493715859375\\
9	0.0113423490898438\\
10	0.00690336968320313\\
11	0.00417727052931641\\
12	0.0025165756958584\\
13	0.00151086797327788\\
14	0.00090458241588148\\
15	0.000540380505387779\\
};

\addplot [color=blue!90!white,line width=0.8pt,only marks,mark=o,mark options={solid,scale=1.25}]
  table[row sep=crcr]{%
2	0.2775\\
3	0.1655\\
};

\addplot [color=white!50!black!50,line width=0.8pt,mark=*,mark options={red, scale=0.5, solid}]
  table[row sep=crcr]{%
2	0.234375\\
3	0.08203125\\
4	0.0439453125\\
5	0.016845703125\\
6	0.00567626953125\\
7	0.0028839111328125\\
8	0.00108718872070313\\
9	0.000363349914550781\\
10	0.000182390213012695\\
11	6.84857368469238e-05\\
12	2.28434801101685e-05\\
13	1.14329159259796e-05\\
14	4.28874045610428e-06\\
15	1.42981298267841e-06\\
};

\addplot [color=red,line width=0.8pt]  table[row sep=crcr]{%
2	0.234375\\
3	0.08203125\\
4	0.0263671875\\
5	0.008056640625\\
6	0.00238037109375\\
7	0.0006866455078125\\
8	0.000194549560546875\\
9	5.43594360351563e-05\\
10	1.50203704833984e-05\\
11	4.11272048950195e-06\\
12	1.11758708953857e-06\\
13	3.01748514175415e-07\\
14	8.10250639915466e-08\\
15	2.16532498598099e-08\\
};

\addplot [color=red,line width=0.8pt,only marks,mark=o,mark options={solid,scale=1.25}]
  table[row sep=crcr]{%
2	0.234375\\
3	0.08203125\\
};

\node[below, right, align=left, inner sep=0mm, text=black]
at (rel axis cs:0.75,0.86) {BSC};

\node[below, right, align=left, inner sep=0mm, text=black]
at (rel axis cs:0.51,0.65) {erasures + errors};

\node[below, right, align=left, inner sep=0mm, text=black]
at (rel axis cs:0.30,0.5) {BEC};

\end{axis}
\end{tikzpicture}%
\vspace{-2mm}\caption{Error probability for $n$ uses of the channel~\refE{erasure-Pyx}, with $q=2$, $M=4$ and BSC: $(\epsilon,\delta)=(0.25, 0)$, erasures and errors: $(\epsilon,\delta)=(0.05, 0.2)$, and BEC: $(\epsilon,\delta)=(0, 0.25)$.}\label{fig:BSC-BEC-bounds-M4}
\end{figure}

The bound in \refE{MDS-optimality} coincides with the converse bound \cite[Th.~38]{Pol10}. As observed in \cite{Pol10}, this lower bound is tight when $\Cc$ is an MDS code. Here this result is recovered via the definition of generalized quasi-perfect codes.

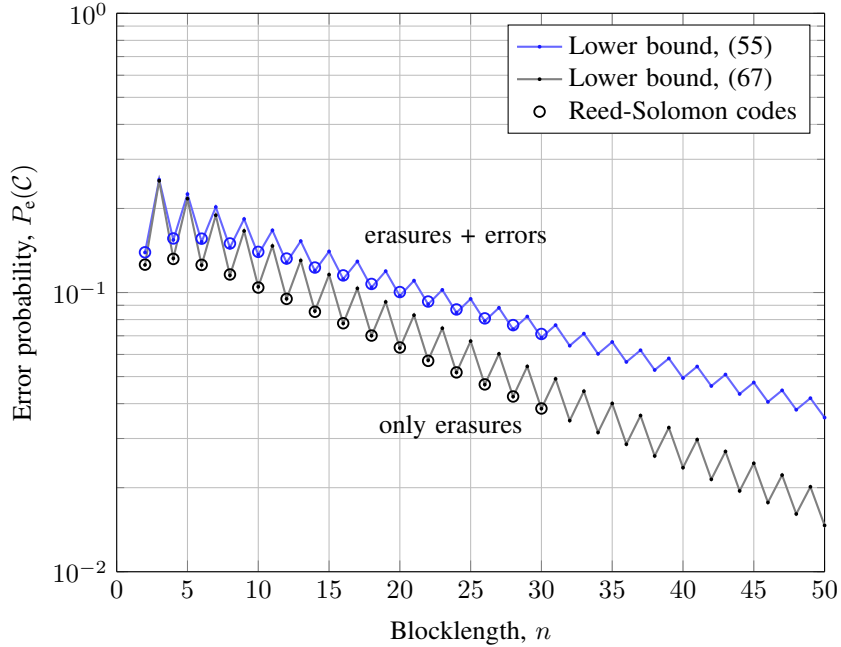
\begin{figure}[t]%
\centering
%
%
\begin{tikzpicture}

\begin{axis}[%
width=0.57\textwidth,
height=0.45\textwidth,
at={(0\textwidth,0\textwidth)},
scale only axis,
separate axis lines,
every outer x axis line/.append style={black},
every x tick label/.append style={font=\color{black}},
xmin=0,
xmax=50,
xlabel={Blocklength, $n$},
xmajorgrids,
every outer y axis line/.append style={black},
every y tick label/.append style={font=\color{black}},
ymode=log,
ymin=0.01,
ymax=1,
yminorticks=true,
ylabel={Error probability, $\Pe(\Cc)$},
ymajorgrids,
yminorgrids,
legend style={legend cell align=left,align=left,fill=white}
]

\addplot [color=blue!60!white,line width=0.8pt,mark=o,mark options={blue!90!white,scale=0.15, solid}]
  table[row sep=crcr]{%
2	0.13915\\
3	0.253672482247032\\
4	0.154328338145161\\
5	0.225181113653985\\
6	0.151989935615875\\
7	0.202408623184678\\
8	0.145000671230299\\
9	0.183373657741637\\
10	0.13664371010843\\
11	0.16706223636155\\
12	0.128059285155632\\
13	0.152859053588808\\
14	0.119698145731109\\
15	0.140348250200688\\
16	0.111743974007148\\
17	0.129230548531713\\
18	0.104264821894306\\
19	0.119281200602369\\
20	0.097275336463788\\
21	0.110326065988674\\
22	0.0907648310842276\\
23	0.102226915884652\\
24	0.084710798354187\\
25	0.0948718851399164\\
26	0.0790857037214567\\
27	0.0881689956891933\\
28	0.0738604858007909\\
29	0.082041609231825\\
30	0.0690063677458772\\
31	0.0764251426549101\\
32	0.064495773247471\\
33	0.0712646384245378\\
34	0.0603027572269158\\
35	0.0665129305718104\\
36	0.0564031706170593\\
37	0.0621292358092947\\
38	0.0527746798249518\\
39	0.0580780545656185\\
40	0.0493967084985595\\
41	0.0543283021548447\\
42	0.0462503400049335\\
43	0.0508526136414897\\
44	0.0433182025413833\\
45	0.0476267817155276\\
46	0.0405843493292979\\
47	0.0446292977502919\\
48	0.0380341408262664\\
49	0.04184097384386\\
50	0.0356541326565761\\
51	0.0392446290946764\\
52	0.0334319710577799\\
53	0.0368248273169964\\
54	0.0313562965288424\\
55	0.0345676563119672\\
56	0.0294166557258746\\
57	0.0324605409794197\\
58	0.0276034212961223\\
59	0.0304920841913809\\
60	0.0259077191565881\\
61	0.0286519305955484\\
62	0.0243213626415646\\
63	0.0269306494773486\\
64	0.0228367929215812\\
};
\addlegendentry{Lower bound, \refE{erasure-Q-optimality}};

\addplot [color=black!50!white,line width=0.8pt,mark=*,mark options={black, scale=0.15, solid}]
  table[row sep=crcr]{%
2	0.12555\\
3	0.251242560934205\\
4	0.1324866375\\
5	0.216816792226812\\
6	0.125343796993875\\
7	0.189120393561224\\
8	0.115225770206657\\
9	0.166173347039298\\
10	0.104809498643462\\
11	0.14679965752725\\
12	0.0949154476467378\\
13	0.130226523408398\\
14	0.0857971552833115\\
15	0.115910946038655\\
16	0.07750487988593\\
17	0.103452859326527\\
18	0.0700108993345125\\
19	0.0925469493621699\\
20	0.0632582875069899\\
21	0.0829538416292626\\
22	0.057181323736761\\
23	0.0744818353642887\\
24	0.0517142983182403\\
25	0.0669747718368078\\
26	0.0467952401239729\\
27	0.0603036637700579\\
28	0.0423673254713698\\
29	0.0543607337803563\\
30	0.0383792084373045\\
31	0.0490550533386763\\
32	0.0347848553274465\\
33	0.0443092787962024\\
34	0.0315431651829821\\
35	0.0400571598893672\\
36	0.0286175145245984\\
37	0.0362416050717052\\
38	0.0259752936607821\\
39	0.0328131565715377\\
40	0.0235874661540705\\
41	0.0297287724749877\\
42	0.0214281648250698\\
43	0.0269508426341449\\
44	0.0194743284161332\\
45	0.0244463852513313\\
46	0.017705378437974\\
47	0.0221863849000597\\
48	0.0161029335708698\\
49	0.0201452425720726\\
50	0.0146505581250371\\
51	0.0183003154022843\\
52	0.0133335408623165\\
53	0.0166315288762067\\
54	0.0121387006101583\\
55	0.0151210481368248\\
56	0.0110542153795351\\
57	0.0137529978650891\\
58	0.0100694720350698\\
59	0.0125132223750178\\
60	0.00917493390853418\\
61	0.0113890792258168\\
62	0.00836202407126124\\
63	0.0103692609403114\\
64	0.00762302227609\\
};
\addlegendentry{Lower bound, \refE{MDS-optimality}};

\addplot [color=black,line width=0.7pt,only marks,mark=o,mark options={solid}]
  table[row sep=crcr]{%
2	0.125871\\
4	0.131928\\
6	0.125613\\
8	0.116036\\
10	0.104189\\
12	0.094898\\
14	0.085337\\
16	0.077618\\
18	0.070129\\
20	0.063542\\
22	0.056927\\
24	0.051751\\
26	0.046874\\
28	0.042428\\
30	0.038427\\
};
\addlegendentry{Reed-Solomon codes};

\addplot [color=blue!90!white,line width=0.7pt,only marks,mark=o,mark options={solid}]
  table[row sep=crcr]{%
2	0.139325\\
4	0.156203\\
6	0.155943\\
8	0.14991\\
10	0.139841\\
12	0.132338\\
14	0.122881\\
16	0.115169\\
18	0.107334\\
20	0.100374\\
22	0.09285\\
24	0.086971\\
26	0.080794\\
28	0.076438\\
30	0.071057\\
};

\node[below, right, align=left, inner sep=0mm, text=black]
at (rel axis cs:0.35,0.6) {erasures + errors};

\node[below, right, align=left, inner sep=0mm, text=black]
at (rel axis cs:0.37,0.26) {only erasures};

\end{axis}
\end{tikzpicture}%
\vspace{-6mm}\caption{Error probability for $n$ uses of the $q$-ary channel \refE{erasure-Pyx} with $q=32$, fixed transmission rate $R=\frac{1}{n}\log_q M=\frac{1}{2}$, and  erasures and errors: $(\epsilon,\delta)=(0.05,0.25)$, only erasures: $(\epsilon,\delta)=(0,0.25)$. }\label{fig:qary-bounds-Q32}
\end{figure}

As an example, let us consider the transmission of $M=4$ codewords over a length-$n$ binary input channel~\refE{erasure-Pyx} for three sets of parameters: BSC with $(\epsilon,\delta)=(0.25, 0)$, a channel with erasures and errors with $(\epsilon,\delta)=(0.05, 0.2)$ and BEC with $(\epsilon,\delta)=(0, 0.25)$. \refFig{BSC-BEC-bounds-M4} depicts the exact error probability $\Pe(\Cc)$ of the best code compared with the lower bound~\refE{erasure-Q-optimality} with the choice of $\Psi(e)$ given in~\refE{binary-De}. The optimum codes for the BSC and BEC are taken from \cite{moser2013} and \cite{moser2017}, respectively. For the channel with combined erasures and errors we also use the code for the BEC, since it offers a better performance at  points where it differs from that of the BSC. \refFig{BSC-BEC-bounds-M4} shows that the bound \refE{erasure-Q-optimality} for the BSC coincides with the code error probability at the points were quasi-perfect codes exist with respect to the Hamming distance ($n=2,3,4,5,6,8$). For the BEC, the bound \refE{erasure-Q-optimality} (which coincides with \refE{MDS-optimality}) provides the exact error probability at the points where (trivial) MDS codes exist ($n=2,3$), as they are generalized quasi-perfect. For the combined errors-erasures channel, to be optimal the codes need to be generalized quasi-perfect for both the BSC and BEC, which only occurs at $n=2,3$.

Let us consider now a $q$-ary channel \refE{erasure-pyx} with $q=32$, and fixed transmission rate $R=\frac{1}{n}\log_q M=\frac{1}{2}$. \refF{qary-bounds-Q32} depicts the lower bound \refE{erasure-Q-optimality} (optimized over a family of functions $\Psi(e)$) for combined erasures and errors with $(\epsilon,\delta)=(0.05,0.25)$, and the lower bound \refE{MDS-optimality} for erasures only with $(\epsilon,\delta)=(0,0.25)$. For even blocklengths, we have simulated the performance of a Reed-Solomon code in both scenarios with $10^6$ Monte Carlo realizations. Reed-Solomon codes are defined for blocklengths $n \leq q-1$ and they are generalized quasi-perfect for the $q$-ary erasure channel. Therefore, they attain the lower bound \refE{MDS-optimality} with equality in the erasure-only case. While their performance with errors and erasures is not far from the lower bound \refE{erasure-Q-optimality}, a gap exists in this case. Reed-Solomon codes can be extended for blocklengths $n = q$ and $n=q+1$. There exist no MDS codes for longer blocklengths in general~\cite{seroussi86}.

%

\section{Almost-Lossless Source-Channel Coding}\label{sec:jscc}

In this section, the notion of quasi-perfect codes is generalized to allow non-equiprobable messages, hence the code needs be matched both to the source and the channel. 

We consider the almost-lossless source-channel coding setting. A source generates messages $v\in\Vc$, where $\Vc$ is a discrete alphabet, according to $\pv$. The message $v$ is to be transmitted over a channel $\pyx$, $x\in \Xc$ and $y\in \Yc$, using a channel encoder that maps each source message $v$ into a codeword $x_v \in \Xc$. We let $\pxvC$ denote the conditional distribution $\pxv$ induced by the codebook $\Cc \triangleq \bigl\{ x_1 ,\ldots, x_{|\Vc|} \bigr\}$.
The receiver uses maximum-a-posteriori (MAP) decoding to decide on the transmitted message $\hat v\in\Vc$. This decoder minimizes the average error probability, which is given by
\begin{align}
  \Pe(\Cc) &= \PP\bigl[\hat V \neq V\bigr]\\
    &= 1 - \sum_{y} \max_{v} \pv(v)\pyx\bigl(y|x_v\bigr).
    \label{eqn:jscc-errorMAP}
\end{align}

The concept of generalized quasi-perfect codes presented in Section~\ref{sec:generalized-quasi-perfect} can be extended to almost-lossless source-channel coding.

\begin{definition}\label{def:jscc-perfect-Q-code}
A source-channel code $\Cc$ is \textit{generalized perfect} with respect to a given source $\pv$ and channel $\pyx$, if there exists $\gamma \geq 0$ and an auxiliary distribution $\qy\in\Qc$ such that
\begin{align}\label{eqn:jscc-cup-codes}
 \bigcup_{v\in\Vc} \Sph{x_v}\left(\frac{\gamma}{\pv(v)}, \qy\right) = \Yc,
\end{align}
where the union is disjoint.
More generally, a code is \textit{generalized quasi-perfect} if  there exists $\gamma \geq 0$ and $\qy\in\Qc$ such that \refE{jscc-cup-codes} is satisfied and the codeword-centered sets $\left\{ \Sphi{x_v}\Bigl(\frac{\gamma}{\pv(v)},\qy\Bigr), v \in \Vc\right\}$ are disjoint.
\end{definition}

The definition of a source-channel quasi-perfect code induces a packing of spheres with radius depending on the probability of the associated source message -- more probable source messages are associated to larger spheres. 
If the source messages are equiprobable, then the radius of the spheres becomes independent of the associated source message and Definition~\ref{def:jscc-perfect-Q-code} boils down to Definition~\ref{def:perfect-Q-code}.

\begin{theorem}\label{thm:jscc-optimality}
Let $\pv$ be the distribution of the source messages, $\pyx$ be a symmetric channel according to Definition \ref{def:symmetric-DMC}, and $\Cc$ be a generalized quasi-perfect source-channel code according to Definition \ref{def:jscc-perfect-Q-code}. Then,
\begin{align}
  \Pe(\Cc)
    &= \min_{\pxv} \max_{\qy\in\Qc} \left\{
                 \alpha_{\frac{1}{|\Vc|}} \Bigl(\pv \times\pxv \times\pyx,\;
                 \barpv\times \pxv \times \qy \Bigr) \right\}, \label{eqn:jscc-optimality}
\end{align}
where $\barpv(v)=\frac{1}{|\Vc|}$ for all $v\in\Vc$.
Conversely, if \refE{jscc-optimality} holds, then 
$\Cc$ is generalized $Q$-quasi-perfect with respect to the source $\pv$ and channel $\pyx$.
\end{theorem}
\begin{IEEEproof}
See Appendix \ref{sec:jscc-optimality}.
\end{IEEEproof}

The right-hand side of \refE{jscc-optimality} is precisely  the converse bound \cite[Th. 4]{Kost13} particularized in the almost-lossless setting. Therefore, Theorem \ref{thm:jscc-optimality} shows that \cite[Th. 4]{Kost13} is tight provided that a generalized quasi-perfect code matched to the source and channel exists.

As a particular case, consider a noiseless channel such that $y=x$ with $\Xc = \Yc = \{1,\ldots,M\}$, and $|\Vc|>M$. In this case, Definition \ref{def:jscc-perfect-Q-code} yields ``spheres'' of size $1$ for the $M$ most probable messages and the $|\Vc|-M$ least probable messages are assigned to ``empty spheres''. In practice, the messages associated to these ``empty spheres'' can be assigned to an arbitrary channel index, as they always yield to a decoding error given their smaller probability. This code corresponds precisely to the well-known optimal almost-lossless block source code.
When the $M$ most probable messages have a strictly larger probability than that of the $|\Vc|-M$ least probable messages, the code is generalized perfect according to Definition~\ref{def:jscc-perfect-Q-code}. When the $M$-th and $(M+1)$-th most probable messages have the same probability, the code is generalized quasi-perfect.

\section{Lossy Source Coding}\label{sec:lsc}

In this section, we consider the lossy source coding problem with a maximum distortion constraint. A source generates messages $v\in\Vc$ with probability distribution $\pv$. The source encoder maps the message $v$ to a codeword $w\in\Wc$ belonging to a length-$M$ codebook $\Cc = \{w_1, w_2, \ldots, w_M\}$. Here $\Wc$ denotes the reconstruction alphabet. We define a non-negative real-valued distortion measure $d(v,w): \Vc\times\Wc \to \RR^{+}$ and consider a maximum allowed distortion $D$. The minimum excess-distortion probability of a given code $\Cc$ is defined as
\begin{align}
   \Ped(\Cc,D)
     &\triangleq \PP\bigl[d(V,W) > D\bigr]\\
     &= 1 - \PP\left[ \min_{w\in\Cc} d(V,w) \leq D \right], \label{eqn:P-excess-distortion}
\end{align}
where in \refE{P-excess-distortion} we used that the minimum excess distortion probability is attained by assigning each source message to the closest (in terms of distortion measure) codeword $w\in\Cc$.

Quasi-perfect codes have good packing and covering properties simultaneously. Therefore, they are both good channel, as shown in the previous sections, and source codes, as shown next. According to Definition~\ref{def:perfect-Q-code} whether a code is generalized quasi-perfect code depends on the channel. In the lossy source-coding setting, this channel corresponds to the test channel induced by the rate-distortion function.

Consider a block source encoder that encodes $n$ realizations of the source $\pv$ using a codebook of cardinality~$2^{nR}$. Rate-distortion theory states that, as the blocklength $n$ grows large, the largest rate $R$ of a codebook with maximum distortion $D$ and vanishing excess-distortion probability is given by
\begin{align}
   R(D) \triangleq \min_{\pwv: \Ex[d(V,W)\leq D]} I(V;W).
   \label{eqn:RD-def}
\end{align}
The optimal $\pwv^{\star}$ in \refE{RD-def} induces a \textit{test channel} $\pvw^{\star}$ that maps the reconstruction points into the source alphabet.
More precisely, let $\pw^{\star}(w) = \sum_v \pv(v)\pwv^{\star}(w|v)$, then, Bayes' rule yields $\pvw^{\star}(v|w) = \frac{\pv(v)\pwv^{\star}(w|v)}{\pw^{\star}(w)}$.
It is shown in~\cite[Sec. 10.7]{Cover06} that the optimal test channel has the form 
\begin{align}
   \pvw^{\star}(v|w) 
     &= \frac{\pv(v)e^{-\lambda^{\star} d(v,w)}}{\mu(v)},
     \label{eqn:RD-test-channel}
\end{align}
for some $\lambda^{\star}\geq 0$, and where $\mu(v) = \sum_w \pw^{\star}(w) e^{-\lambda^{\star} d(v,w)}$ is a normalization constant independent of $w$.

Let us consider the channel coding problem, as described in \refS{generalized-quasi-perfect}, of transmitting $M$ messages over the channel $\pvw^{\star}$. Good channel codes for $\pvw^{\star}$ become good source codes for the source $\pv$ and distortion measure $d(v,w)$. In particular, quasi-perfect codes attain the minimum excess-distortion probability, as the next result shows.

\begin{theorem}\label{thm:lsc-optimality}
Consider a source $\pv$ with $\pv(v)>0$, $v\in\Vc$, distortion measure $d(v,w)$ and maximum distortion~$D$. Let the test channel $\pvw^{\star}$ in \refE{RD-test-channel} be symmetric according to Definition \ref{def:symmetric-DMC} and let $\tilde\qy(v)=\frac{1}{c_{\mu}}\frac{\pv(v)}{\mu(v)}$ satisfy $\tilde\qy\in\Qc$,  where $\mu(v)$ is the normalizing factor in \refE{RD-test-channel} and $c_{\mu}\triangleq\sum_{v'} \frac{\pv(v')}{\mu(v')}$.
Then, for any size-$M$ code $\Cc$, generalized quasi-perfect according to Definition \ref{def:perfect-Q-code} with parameters $\gamma$ and $\qy=\tilde\qy$, it holds that
\begin{align}\label{eqn:lsc-optimality}
   \Ped(\Cc,D) =  \max\limits_{\qv}\Bigl\{\alpha_{M \sup_{w\in\Cc} \QQ_V\left[d(V,w) \leq D \right]}\bigl(\pv, \qv\bigr) \Bigr\},
\end{align}
where $\QQ_{V}[\cdot]$ denotes the probability when $V\sim \qv$.

Moreover, if $D \geq - \frac{1}{\lambda^{\star}} \log\bigl( \gamma/c_{\mu} \bigr)$, the excess-distortion probability is $\Ped(\Cc,D) = 0$.
\end{theorem}
\begin{IEEEproof}
See Appendix \ref{sec:lsc-optimality}.
\end{IEEEproof}

In \cite[Th. 3]{vazquez16}, the excess-distortion probability of any source code $\Cc$ (not necessarily quasi-perfect) is expressed as the error probability of an induced binary hypothesis test with certain parameters,
\begin{align}
  \Ped(\Cc,D)
    &= \max_{\qv}
      \Bigl\{\alpha_{\QQ_{V}[ d(V,\Cc) \leq D ]}
         \bigl(\pv, \qv\bigr) \Bigr\},
    \label{eqn:lsc-tight}
\end{align}
where $d(v,\Cc) \triangleq \min_{w\in\Cc} d(v,w)$. 
Invoking
\begin{align}
\QQ_{V}[ d(V,\Cc) \leq D ] 
\ \leq\ M \sup_{w\in\Cc} \QQ_V\left[d(V,w) \leq D \right]
\ \leq\ M \sup_{w\in\Wc} \QQ_V\left[d(V,w) \leq D \right],
\end{align}
the identity \refE{lsc-tight} yields the lower bounds
\begin{align}
  \Ped(\Cc,D) &\geq \max_{\qv}\Bigl\{\alpha_{M \sup_{w\in\Cc} \QQ_V\left[d(V,w) \leq D \right]}\bigl(\pv, \qv\bigr) \Bigr\}
    \label{eqn:lsc-bound}\\
    &\geq \max_{\qv}
      \Bigl\{\alpha_{M \sup_{w\in\Wc} \QQ_V\left[d(V,w) \leq D \right]} \bigl(\pv, \qv\bigr) \Bigr\}.
    \label{eqn:lsc-bound-kostinathm8}
\end{align}
Theorem \ref{thm:lsc-optimality} shows that, provided that a quasi-perfect code exists with certain parameters, the lower bound \refE{lsc-bound} holds with equality. The relaxation over the reconstruction alphabet in \refE{lsc-bound-kostinathm8} coincides with \cite[Th.~8]{Kost12}. For certain sources, the inequality \refE{lsc-bound-kostinathm8} may hold with equality as the next example shows.

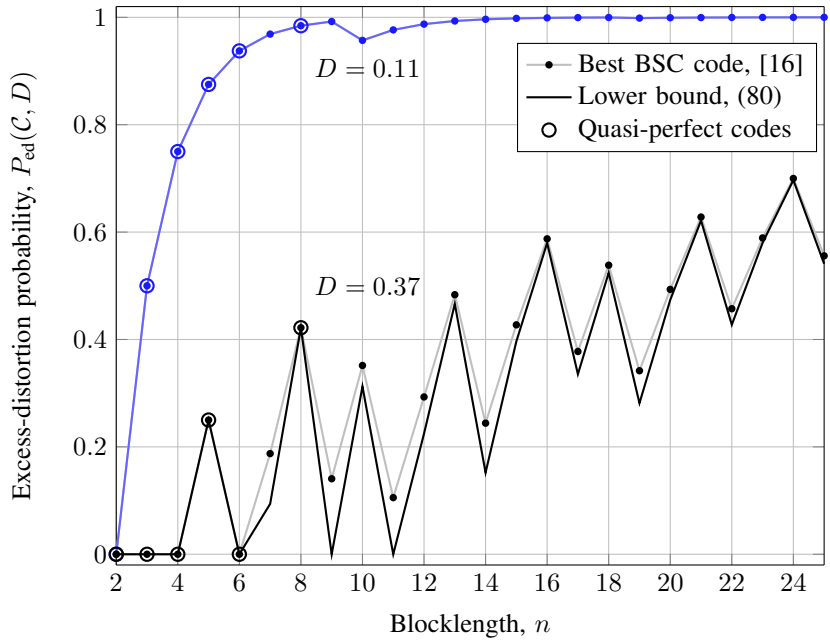
\begin{figure}[t]
    \centering
%
%
\begin{tikzpicture}

\begin{axis}[%
width=0.57\textwidth,
height=0.45\textwidth,
at={(0\textwidth,0\textwidth)},
scale only axis,
separate axis lines,
every outer x axis line/.append style={black},
every x tick label/.append style={font=\color{black}},
xmin=2,
xmax=25,
xmajorgrids,
xlabel={Blocklength, $n$},
every outer y axis line/.append style={black},
every y tick label/.append style={font=\color{black}},
ymin=-0.02,
ymax=1.02,
ymajorgrids,
ylabel={Excess-distortion probability, $\Ped(\Cc,D)$},
legend style={at={(0.99,0.935)},anchor=north east,legend cell align=left,align=left,draw=black}
]

\addplot [color=white!50!black!50,line width=0.8pt,mark=*,mark options={blue!90!white, scale=0.5, solid}, forget plot]
  table[row sep=crcr]{%
2	0\\
3	0.5\\
4	0.75\\
5	0.875\\
6	0.9375\\
7	0.96875\\
8	0.984375\\
9	0.9921875\\
10	0.95703125\\
11	0.9765625\\
12	0.9873046875\\
13	0.9931640625\\
14	0.996337890625\\
15	0.998046875\\
16	0.99896240234375\\
17	0.99945068359375\\
18	0.999710083007813\\
19	0.998542785644531\\
20	0.999195098876953\\
21	0.999557495117188\\
22	0.999757766723633\\
23	0.999867916107178\\
24	0.99992823600769\\
25	0.999961137771606\\
};

\addplot [color=blue!60!white,solid,line width=0.8pt, forget plot]
  table[row sep=crcr]{%
2	0\\
3	0.5\\
4	0.75\\
5	0.875\\
6	0.9375\\
7	0.96875\\
8	0.984375\\
9	0.9921875\\
10	0.95703125\\
11	0.9765625\\
12	0.9873046875\\
13	0.9931640625\\
14	0.996337890625\\
15	0.998046875\\
16	0.99896240234375\\
17	0.99945068359375\\
18	0.999710083007813\\
19	0.998542785644531\\
20	0.999195098876953\\
21	0.999557495117188\\
22	0.999757766723633\\
23	0.999867916107178\\
24	0.99992823600769\\
25	0.999961137771606\\
};

\addplot [color=blue!90!white,line width=0.8pt,only marks,mark=o,mark options={solid,scale=1.25}, forget plot]
  table[row sep=crcr]{%
2	0\\
3	0.5\\
4	0.75\\
5	0.875\\
6	0.9375\\
8	0.984375\\
};

\addplot [color=white!50!black!50,line width=0.8pt,mark=*,mark options={black, scale=0.5, solid}]
  table[row sep=crcr]{%
2	0\\
3	0\\
4	0\\
5	0.25\\
6	0\\
7	0.1875\\
8	0.421875\\
9	0.140625\\
10	0.3515625\\
11	0.10546875\\
12	0.29296875\\
13	0.4833984375\\
14	0.244140625\\
15	0.42724609375\\
16	0.58746337890625\\
17	0.377655029296875\\
18	0.538330078125\\
19	0.341796875\\
20	0.493282318115234\\
21	0.6280517578125\\
22	0.457300186157227\\
23	0.589125633239746\\
24	0.699889183044434\\
25	0.555783748626709\\
};
\addlegendentry{Best BSC code,~\cite{moser2013}};

\addplot [color=black,line width=0.8pt]
  table[row sep=crcr]{%
2	0\\
3	0\\
4	0\\
5	0.25\\
6	0\\
7	0.09375\\
8	0.421875\\
9	0\\
10	0.3125\\
11	0\\
12	0.224609375\\
13	0.46630859375\\
14	0.152099609375\\
15	0.396484375\\
16	0.57977294921875\\
17	0.33538818359375\\
18	0.52423095703125\\
19	0.28143310546875\\
20	0.473648071289063\\
21	0.621505737304688\\
22	0.42744255065918\\
23	0.579920768737793\\
24	0.696820735931396\\
25	0.540954113006592\\
};
\addlegendentry{Lower bound, \eqref{eqn:lsc-bound-kostinathm8}};

\addplot [color=black,line width=0.8pt,only marks,mark=o,mark options={solid,scale=1.25}]
  table[row sep=crcr]{%
2	0\\
3	0\\
4	0\\
5	0.25\\
6	0\\
8	0.421875\\
};
\addlegendentry{Quasi-perfect codes};

\node[below, right, align=left, inner sep=0mm, text=black]
at (rel axis cs:0.28,0.89) {$D=0.11$};

\node[below, right, align=left, inner sep=0mm, text=black]
at (rel axis cs:0.28,0.5) {$D=0.37$};

\end{axis}
\end{tikzpicture}%
	\vspace{-2mm}\caption{Minimum excess-distortion probability for $n$ i.i.d. samples of a equiprobable BMS with bit error rate distortion measure and parameters $M=4$, and $D=0.11$ (in blue) and $D=0.37$ (in black).}\label{fig:EBMS-bounds-M4}
\vspace{-2mm}
\end{figure}

Let us consider the lossy compression of $n$ i.i.d. samples of an equiprobable binary memoryless source (BMS) with bit error rate distortion measure, i.e., $\Pv(\v)=2^{-n}$ and $d(\v,\w) = \frac{1}{n} \sum_{i=1}^n \openone[v_i\neq w_i]$, with $\v,\w \in \{0,1\}^n$. The test channel for this rate-distortion problem corresponds to a BSC with certain parameters.
As in the channel coding example from \refF{BSC-BEC-bounds-M4}, we consider a codebook with $M=4$ codewords. \refFig{EBMS-bounds-M4} depicts the minimum excess-distortion probability $\Ped(\Cc,D)$ as a function of $n$ for a maximum distortion $D=0.11$ and $D=0.37$. Since we are ``quantizing'' a space of increasing dimension $n$ with only $M=4$ codewords, the excess-distortion probability tends to $1$ as $n\to\infty$ for any $D<\frac{1}{2}$. In \refF{EBMS-bounds-M4}, we plot the lower bound \refE{lsc-bound-kostinathm8} evaluated for $\qv$ uniform \cite[Th.~15]{Kost12}, compared to the exact excess-distortion probability evaluated for the best code in a BSC channel and $M=4$ codewords~\cite{moser2013}. We also highlight with markers the points where quasi-perfect codes exist for the BSC channel, corresponding to $n=2,3,4,5,6,8$ (see \refF{BSC-BEC-bounds-M4}).

In \refF{EBMS-bounds-M4}, we observe that the exact excess-distortion probability coincides with the lower bound \refE{lsc-bound-kostinathm8} at the points where quasi-perfect codes exist both for $D=0.11$ and $D=0.37$. Nevertheless, in the lossy compression setting, the reverse implication is not always true. Depending on the system parameters, the exact excess-distortion probability and the lower bound can also coincide when no quasi-perfect code exist for the corresponding test channel. Indeed, for $D=0.37$ the only points where the exact excess-distortion probability and the lower bound coincide are only those in which quasi-perfect codes exist, while for $D=0.11$ these two expressions coincide for all values of $n$, regardless of whether the code is quasi-perfect. This occurs when the sets $\{ v\in \Vc \,|\, d(V,w) \leq D\}$, $w\in\Cc$, are non-overlapping (this occurs in our example for $D$ sufficiently small). Then, the encoding regions which satisfy the maximum distortion cap are ``spheres'' regardless the specific structure of the
codebook $\Cc$ and the lower bound \refE{lsc-bound-kostinathm8}  yields the exact excess-distortion probability.




%
%
%
%

\section{Discussion}\label{sec:discussion}

We have proposed a generalization of perfect and quasi-perfect codes beyond the Hamming distance and their conventional application to binary symmetric channels. 
The definition of these codes follows from
the packing and covering properties of a set of ``spheres.'' Since the shape of these spheres depends on the channel considered, quasi-perfect codes can only be defined with respect to a specific channel. For the BSC, quasi-perfect codes are defined with respect to the Hamming distance and our definition recovers the classical definition of quasi-perfect codes in the coding literature. Our approach capitalizes on the fact that the shape of these spheres can be tilted by using an auxiliary measure. This allows to extend the proposed formulation to encompass MDS codes, which are shown to be quasi-perfect for erasure channels.

While the proofs of the results in this paper are presented for discrete channels, they can be readily extended for channels with continuous outputs.
In fact, Lemma~\ref{lem:HT-im-formulation}, Definition~\ref{def:symmetric-DMC}, and Theorem~\ref{thm:cc-Q-optimality} apply without change for both discrete and continuous channels, provided that they are absolutely continuous with respect to the Lebesgue measure. Nevertheless, the spheres induced by typical continuous channels seldom allow a perfect (or quasi-perfect) packing of the output space. 
Some atypical examples of continuous channels in which the induced spheres pack the output space are the AWGN channel with $M=2$ codewords, the binary-input AWGN when all the input sequences are used, i.e., for $M=2^n$, or the additive white Laplace noise channel (as the induced spheres are norm-$1$ balls, and thus they can pack the space for specific lattice codes).

The framework presented in this work has been built upon the assumption that certain channel symmetry exists. Nevertheless, the underlying idea can be applied to general channels $\pyx$ and arbitrary auxiliary distributions $Q$. In this case, quasi-perfect codes are defined as those ``codes attaining the meta-converse bound with equality.'' This definition is reminiscent to that of the MDS codes, which are defined as ``codes attaining the Singleton bound with equality.'' While this alternative more general definition of quasi-perfect codes is mathematically precise, it does not shed much light into the structure of the corresponding quasi-perfect codes. 

\appendices
\section{Proof of Theorem \ref{thm:jscc-optimality}}\label{sec:jscc-optimality}

We apply Lemma \ref{lem:HT-im-formulation} to the hypothesis test in~\refE{jscc-optimality} to obtain an alternative expression for the Neyman-Paerson performance of the test. This expression is then shown to coincide with the following characterization of the joint source-channel error probability of a quasi-perfect code.

\begin{lemma}\label{lem:jscc-Q-error}
For a source $\pv$ and a symmetric channel $\pyx$, let the source-channel code $\Cc$ be generalized quasi-perfect with parameters $\gamma$ and $\qy$ according to Definition \ref{def:jscc-perfect-Q-code}. The error probability of $\Cc$ is
\begin{align}
  \Pe(\Cc)
     &= \sum_{v} \sum_{\frac{\tau}{\pv(v)} \in \Lc_{\qy},\, \tau\leq \gamma} \tau \mathsf{Q}_{\text{o}}\left(\tfrac{\tau}{\pv(v)}\right)
   - \gamma \biggl(1 - \sum_{v} \mathsf{Q}_{\text{i}}\left(\tfrac{\gamma}{\pv(v)}\biggr) \right).
    \label{eqn:jscc-Q-error}
\end{align}
\end{lemma}
\begin{IEEEproof}
The proof follows exactly the same steps as that of Lemma \ref{lem:cc-Q-error}, and it is omitted here. \end{IEEEproof}

Applying Lemma \ref{lem:HT-im-formulation} with $P_0 \leftarrow \pv \pxv \pyx$ and $P_1 \leftarrow \barpv \pxv \qy$, via the change of variable $\gamma \leftrightarrow \gamma' = \frac{\gamma}{|\Vc|}$,  yields
\begin{align}
\alpha_{\frac{1}{|\Vc|}} \Bigl(\pv \pxv \pyx,\,              \barpv \pxv \qy \Bigr)
&= \max_{\gamma' \geq 0} \bigggl\{ \sum_{v,x}\pv(v) \pxv(x|v) \sum_{y\notin \Sphi{x}\left(\frac{\gamma'}{\pv(v)},\qy\right)} \frac{\pyx(y|x)}{\qy(y)} \qy(y) \notag\\ 
&\qquad\qquad      +\, \gamma' \sum_{v,x} \pxv(x|v) \sum_{y\in \Sphi{x}\left(\frac{\gamma'}{\pv(v)},\qy\right)} \qy(y) 
           \,-\, \gamma' \bigggr\}
           \label{eqn:jscc-Q-alpha-2}\\
  &= \max_{\gamma' \geq 0} \bigggl\{ \sum_{v,x}\pv(v) \pxv(x|v)
\sum_{\tau \in \Lc_Q, \tau\leq\frac{\gamma'}{\pv(v)}} \tau \mathsf{Q}_{\text{o}}(\tau)\notag\\ 
&\qquad\qquad  +\, \gamma' \sum_{v,x} \pxv(x|v) 
 \,\mathsf{Q}_{\text{i}}\left(\tfrac{\gamma'}{\pv(v)}\right) \,-\, \gamma' \bigggr\},
           \label{eqn:jscc-Q-alpha-3}
\end{align}
where in the last step we used that the complementary set of $\Sphi{x}(\gamma,,\qy)$ corresponds to $\bigcup_{\tau \in \Lc, \tau \leq \gamma} \Spho{x}(\tau,\qy)$ and that  $\frac{\pyx(y|x)}{\qy(y)}=\tau$ for all $y \in \Spho{x}(\tau,\qy)$.
Finally, using that $\sum_{x} \pxv(x|v) = 1$, and applying the change of variable $\tau' = \tau \pv(v)$ (or equiv. $\tau =  \frac{\tau'}{\pv(v)}$) for each $v$ we obtain
\begin{align}
\alpha_{\frac{1}{|\Vc|}} \Bigl(\pv \pxv \pyx,\,              \barpv \pxv \qy \Bigr)
&= \max_{\gamma' \geq 0} \bigggl\{ \sum_{v}
\sum_{\frac{\tau'}{\pv(v)} \in \Lc_Q, \tau'\leq\gamma'} \tau' \mathsf{Q}_{\text{o}}\left(\tfrac{\tau'}{\pv(v)}\right) \,+\, \gamma' \sum_{v} 
 \,\mathsf{Q}_{\text{i}}\left(\tfrac{\gamma'}{\pv(v)}\right) \,-\, \gamma' \bigggr\},
           \label{eqn:jscc-Q-alpha-4}
\end{align}
which coincides with \refE{jscc-Q-error} when $\gamma$ is the optimizing value of $\gamma'$ in \refE{jscc-Q-alpha-4}. Since \refE{jscc-Q-alpha-4} is a lower bound to $\Pe(\Cc)$, the theorem thus follows by optimizing \refE{jscc-Q-alpha-4} over auxiliary distributions $\qy\in\Qc$.

\section{Proof of Theorem \ref{thm:lsc-optimality}} \label{sec:lsc-optimality}

Let $\Cc$ be generalized quasi-perfect with respect to the test channel $\pvw^{\star}$ defined in \refE{RD-test-channel}, with parameters $\gamma$ and $\tilde\qy(v)=\frac{1}{c_{\mu}}\frac{\pv(v)}{\mu(v)}$. The set $\Sph{w}(\tau,\tilde\qy)$ associated to the test channel $\pwv^{\star}$ is given by
\begin{align}
  \Sph{w}(\tau,\tilde\qy) 
    &= \left\{ v \in \Vc \,\Big|\, d(v,w) \leq - \frac{1}{\lambda^{\star}} \log\left( \tau  \frac{\mu(v) \tilde\qy(v)}{\pv(v)} \right) \right\},\label{eqn:B-lsc}
\end{align}
which upon particularization to $\tilde\qy(v)=\frac{1}{c_{\mu}}\frac{\pv(v)}{\mu(v)}$ yields
\begin{align}
  \Sph{w}(\tau,\tilde\qy) 
    &= \left\{ v \in \Vc \,\Big|\, d(v,w) \leq - \frac{1}{\lambda^{\star}} \log\left( {\tau}/{c_{\mu}}\right) \right\}.\label{eqn:B-lsc-bis}
\end{align}

We  divide the proof in two different cases depending on the value of the maximum distortion $D$. 
\begin{enumerate}
\item $D \geq - \frac{1}{\lambda^{\star}} \log\bigl( \gamma/c_{\mu} \bigr)$. In this case $\gamma \geq c_{\mu} e^{-\lambda^{\star} D}$, and
\begin{align}
\Sph{w}(\gamma,\tilde\qy) &\subseteq   \Sph{w}\bigl(c_{\mu} e^{-\lambda^{\star} D},\tilde\qy\bigr)
 = \Bigl\{ v \in \Vc \,\big|\, d(v,w) \leq D \Bigr\}.
 \label{eqn:Bgamma-subsetBD-1}
\end{align}
According to Definition \ref{def:perfect-Q-code},
the codeword-centered sets $\Sph{w}(\gamma,\tilde\qy)$, $w \in \Cc$, cover the space. Then, using \refE{Bgamma-subsetBD-1} it follows that 
\begin{align}
 \bigcup_{w\in\Cc} 
 \Bigl\{ v \in \Vc \,\big|\, d(v,w) \leq D \Bigr\} = \Vc.\label{eqn:cup-lsc-codes}
\end{align}
As a result, the excess-distortion probability is
\begin{align}
   \Ped(\Cc,D)
     &= 1 - \sum_{v} \pv(v) \openone\left[ \min_{w\in\Cc} d(v,w) \leq D \right]\\
     &= 1 - \sum_{v} \pv(v) \openone\biggl[ v \in \medcup_{w\in\Cc}\Bigl\{ v \in \Vc \,\big|\, d(v,w) \leq D \Bigr\} \biggr] \;=\; 0. \label{eqn:cup-lsc-Ped0}
\end{align}
According to \refE{cup-lsc-codes}, for any distribution $\qv$ defined over $\Vc$, we have that $\QQ_V\bigl[ d(V,\Cc) \leq D \bigr] = 1$. This indeed implies that $\sup_{w\in\Cc} \QQ\left[d(V,w) \leq D \right] \geq \frac{1}{M}$. Since $\alpha_{1} \bigl(\pv, \qv\bigr) = 0$, using \refE{cup-lsc-Ped0}, we conclude that \refE{lsc-optimality} holds with equality.

\item $D < - \frac{1}{\lambda^{\star}} \log\bigl( \gamma/{c_{\mu}} \bigr)$. In this region, $\gamma < c_{\mu} e^{-\lambda^{\star} D}$, and it thus follows that
\begin{align}
    \Sphi{w}(\gamma,\tilde\qy) &\supseteq \Sph{w}\bigl(c_{\mu} e^{-\lambda^{\star} D},\tilde\qy\bigr)= \Bigl\{ v \in \Vc \,\big|\, d(v,w) \leq D \Bigr\}.
 \label{eqn:Bgamma-subsetBD-2}
\end{align}
In this case, $\bigcup_{w\in\Cc} \bigl\{ v \in \Vc \,|\, d(v,w) \leq D \bigr\}$  does not cover the space completely. Nevertheless, since the code $\Cc$ is quasi-perfect with radius $\gamma$, the spheres $ \Sphi{w}(\gamma,\qy)$, $w \in \Cc$, are disjoint.
Using \refE{Bgamma-subsetBD-2} we conclude that the sets
$\bigl\{ v \in \Vc \,\big|\, d(v,w) \leq D \bigr\}$, $w \in \Cc$, do not overlap. Therefore,
\begin{align}
   \Ped(\Cc,D)
     &= 1 - \PP_V\biggl[ \min_{w\in\Cc} d(V,w) \leq D \biggr]\\
      &= 1 - \PP_V \biggl[ v \in {\textstyle \bigcup\limits_{w\in\Cc}} \Bigl\{ v \in \Vc \,\big|\, d(v,w) \leq D \Bigr\} \biggr]\\
     &= 1 - \sum_{w\in\Cc} \PP_V \bigl[ d(V,w) \leq D \bigr]. \label{eqn:Ped-case2}
\end{align}

We now show that the right-hand side of \refE{lsc-optimality} coincides with \refE{Ped-case2}. Applying Lemma \ref{lem:HT-im-formulation} to the hypothesis test in~\refE{lsc-optimality}, yields
\begin{align}
\alpha_{\beta}\bigl(\pv, \qv\bigr) &=   \max_{\gamma' \geq 0} \left\{\PP_V\biggl[  \frac{\pv(V)}{\qv(V)} \leq \gamma'\biggr]
      + \gamma' \QQ_V\biggl[ \frac{\pv(V)}{\qv(V)} > \gamma'\biggr]
           - \gamma' \beta \right\}\label{eqn:Ped-alpha-1b}
\end{align}

Let
\begin{align}
  \qv^{\Cc}(v) \triangleq \frac{1}{g} \pv(v) \left( \frac{1}{M} \sum_{w\in\Cc} e^{-\lambda d(v,w)}\right)^{-1}
\end{align}
where $g$ a normalizing factor and $\lambda\geq 0$ is to be defined later. Using $\qv = \qv^{\Cc}$ and choosing $\gamma'= \frac{g}{M} e^{-\lambda D}$, we obtain the following lower bound to \refE{Ped-alpha-1b},
\begin{align}
\alpha_{\beta}\bigl(\pv,\qv^{\Cc}\bigr) \geq
 \PP_V\Biggl[\sum_{w\in\Cc} e^{-\lambda d(v,w)} \leq e^{-\lambda D}\Biggr]
      + \frac{g}{M} e^{-\lambda D} \left( \QQ_V\Biggl[ \sum_{w\in\Cc} e^{-\lambda d(v,w)} > e^{-\lambda D} \Biggr] - \beta\right),\label{eqn:Ped-alpha-2b}
\end{align}
where the probability $\QQ_V[\cdot]$ is computed with respect to $\qv = \qv^{\Cc}$.

For $\lambda\geq 0$ sufficiently large, 
\begin{align}
\sum_{w\in\Cc} e^{-\lambda d(v,w)} > e^{-\lambda D}
\;\Leftrightarrow\; \min_{w\in\Cc} d(v,w) \leq D.
\end{align}
Therefore, in this case \refE{Ped-alpha-2b} becomes
\begin{align}
\alpha_{\beta}\bigl(\pv, \qv^{\Cc}\bigr) \geq \PP_V\Bigl[ \min\limits_{w\in\Cc} d(V,w) > D\Bigr] + \frac{g}{M} e^{-\lambda D} \left( \QQ_V\Bigl[ \min\limits_{w\in\Cc} d(V,w) \leq D \Bigr] - \beta\right).\label{eqn:Ped-alpha-3b}
\end{align}
The symmetry conditions required in the theorem imply that
the measure of the set $\bigl\{ v\in\Vc \,|\,  d(v,w) \leq \delta \bigr\}$ does not depend on $w\in\Wc$ for any $\delta\geq 0$. Then, since the sets $\bigl\{ v\in\Vc \,|\,  d(v,w) \leq \delta \bigr\}$ are non-overlapping,
for $\lambda\geq 0$ sufficiently large we obtain
\begin{align}
\QQ_V\Bigl[ \min\limits_{w\in\Cc} d(V,w) \leq D \Bigr] &=
  \sum_{w\in\Cc} \QQ_V\bigl[ d(V,w) \leq D \bigr]\\
  &= M \sup_{w\in\Cc} \QQ_V\left[d(V,w) \leq D \right],
\end{align}
where in the last step we used that, for $\lambda\geq 0$ sufficiently large, $\qv^{\Cc}(v)$ only depends on the distance to the closest $w\in\Cc$. Then, since the measure of the set $\bigl\{ v\in\Vc \,|\,  d(v,w) = \delta \bigr\}$ does not depend on $w\in\Wc$ for any $\delta\geq 0$, neither does $\QQ_V\left[d(V,w) = \delta \right]$ nor $\QQ_V\left[d(V,w) \leq D \right]$ depend on $w\in\Cc$.

Therefore, for $\beta = M \sup_{w\in\Cc} \QQ\left[d(V,w) \leq D \right]$, \refE{Ped-alpha-3b} becomes
\begin{align}
\alpha_{ M \sup_{w\in\Cc} \QQ\left[d(V,w) \leq D \right]}\bigl(\pv, \qv\bigr) 
\geq 1-\sum_{w\in\Cc}\PP_V\bigl[ d(V,w) \leq D\bigr].\label{eqn:Ped-alpha-4b}
\end{align}
Since the left-hand side of \refE{Ped-alpha-4b} is a lower bound to $\Ped(\Cc,D)$, ans since the right-hand side of \refE{Ped-alpha-4b} coincides with \refE{Ped-case2}, then we conclude that \refE{lsc-optimality} holds with equality.

\begin{remark}
Note that for the choice $\qv = \qv^{\Cc}$, by letting $\lambda\to\infty$, $\QQ_V\bigl[ d(V,w) \leq D \bigr]$ becomes independent of $w\in\Cc$. However, for this choice of $\qv$, the measure $\QQ_V\bigl[ d(V,w) \leq D \bigr]$ still depends on $w\notin\Cc$. Therefore, the proof technique presented here cannot be directly applied when the $\beta$ parameter in \refE{lsc-optimality} is relaxed from  $M \sup_{w\in\Cc} \QQ\left[d(V,w) \leq D \right]$ to  $M \sup_{w\in\Wc} \QQ\left[d(V,w) \leq D \right]$, as discussed in \refE{lsc-bound}-\refE{lsc-bound-kostinathm8}.
\end{remark}

\end{enumerate}

\bibliographystyle{IEEEtran}
\bibliography{bib/references}

\end{document}